# Implementation of a fully nonlinear Hamiltonian Coupled-Mode Theory, and application to solitary wave problems over bathymetry

Ch.E. Papoutsellis ([a]) ([b]), A.G. Charalampopoulos ([a]) ([c]), and G.A. Athanassoulis ([a]) ([d])

christos.papoutsellis@centrale-marseille.fr    alexchar@mit.edu    mathan@central.ntua.gr

(a) *School of Naval Architecture and Marine Engineering, National Technical University of Athens(NTUA), Zografos, Greece*

(b) Current address: *Ecole Centrale Marseille and Institut de Recherche sur les Phénomènes Hors Equilibre (IRPHE), Marseille, France*

(c) Current address: *Department of Mechanical Engineering, Massachusetts Institute of Technology, Cambridge, USA*

(d) *Research Center for High Performance Computing, ITMO University, St. Petersburg, Russian Federation*

## Table of contents

**List of abbreviations**

| | |
|---|---|
| BEM | boundary element method |
| CMS | coupled-mode system |
| DNM | direct numerical methods |
| DtN | Dirichlet-to-Neumann (operator) |
| FD | finite difference |
| FDM | finite difference method |
| FEM | finite element method |
| HCMS | Hamiltonian coupled-mode system |
| HCMT | Hamiltonian Coupled-Mode Theory |
| NLPF | nonlinear potential flow |
| NR | Newton-Raphson (method) |
| RK | Runge-Kutta (method) |
| SGN | Serre-Green-Nagdhi (equations) |
| ZCS | Zakharov and Craig & Sulem |

# Implementation of a fully nonlinear Hamiltonian Coupled-Mode Theory, and application to solitary wave problems over bathymetry

**Abstract**

This paper deals with the implementation of a new, efficient, non-perturbative, Hamiltonian coupled-mode theory (HCMT) for the fully nonlinear, potential flow (NLPF) model of water waves over arbitrary bathymetry Papoutsellis and Athanassoulis (2017) (arxiv.org/abs/1704.03276). Applications considered herein concern the interaction of solitary waves with bottom topographies and vertical walls both in two- and three-dimensional environments. The essential novelty of HCMT is a new representation of the Dirichlet-to-Neumann operator, which is needed to close the Hamiltonian evolution equations. This new representation emerges from the treatment of the substrate kinematical problem by means of exact semi-separation of variables in the instantaneous, irregular, fluid domain, established recently by Athanassoulis & Papoutsellis (2017) (https://doi.org/10.1098/rspa.2017.0017). The HCMT ensures an efficient dimensional reduction of the exact NLFP, being able to treat an arbitrary bathymetry as simply as the flat-bottom case, without domain transformation. A key point for the efficient implementation of the method is the fast and accurate evaluation of the space-time varying coefficients appearing in some of its equations. In this paper, all varying coefficients are calculated analytically, resulting in a refined version of the theory, characterized by improved accuracy at significantly reduced computational time. This improved version of HCMT is first validated against existing experimental results and other computations, and subsequently applied to new solitary wave-bottom interaction problems. The latter include: i) the investigation of a new type of "Bragg scattering" effect, appearing when a solitary wave propagates over a seabed with a sinusoidal patch, and ii) the disintegration, focusing and reflection of a solitary wave moving over a three-dimensional bathymetry consisting of parallel banks and troughs, and impinging on a vertical wall.

# 1. Introduction

The study of water waves in intermediate water depth and the reliable modeling of wave-seabed interaction phenomena is a fundamental component of coastal hydrodynamics. The analysis and design of various engineering applications deployed in the coastal environment, requires the development of fast numerical models that can accurately capture the complex wave dynamics in relatively large domains. A basic, long-standing mathematical model, describing various important wave phenomena, is the nonlinear potential flow model, abbreviated subsequently as NLPF. An efficient (re)formulation and implementation of this model, without any simplifications concerning the nonlinearity and the unknown free surface, remains a challenging and useful task. The importance of such developments can be exemplified in many directions, among which we refer to the following two; as a firm basis for developing more realistic models, including breaking and dissipation effects, and as a tool for studying strong interactions between nonlinear waves (e.g. solitary waves) and varying bathymetry or/and vertical boundaries, inducing severe changes in the wave motion and in the involved space and time scales.

Traditionally, in intermediate depth waters, asymptotic methods have been extensively used, based on simplifying assumptions for the free surface and the seabed. The obtained models are dimensionally reduced and easier to solve numerically, in comparison to the complete NLPF formulation, at the expense of being reasonably accurate only to certain limited regions of the fundamental physical parameters (nonlinearity, bottom slope/deformation, shallowness). The extension of the region of validity of these models and their numerical implementation has been the subject of a great deal of work. Restricting attention to models that take into account an uneven seabed, we mention, for instance, i) the Boussinesq-type models [1], [2], [3], [4], [5], [6], [7], [8], [9], ii) the two layer models [10], [11], and iii) the Green-Nagdhi equations [12], [13], [14]. More details and references can be found in Chapter 7 of [15] and in the review articles [16], [17].

In the last two decades or so, the NLPF formulation has been treated numerically by means of various direct numerical methods (DNM). Important examples are the finite difference method (FDM) [18] and the finite element method (FEM) [19], [20], [21], [22] [23], that require the construction of computational grids covering the entire fluid domain. Another popular strategy is the use of boundary element method (BEM), that transfers the computations on the physical boundaries of the fluid domain [24], [25], [26], [27], [28]. The above direct numerical methods take fully into account nonlinearity and dispersion and are quite flexible in the sense that they can treat steep bathymetries, overturning waves, and bodies. However, their application to large domains and long-time simulation requires excessive computational resources, leaving space for lighter and faster solution methods.

The NLPF model, describing an ideal mechanical system, admits of a Hamiltonian formulation, first noticed by Petrov [29] and Zakharov [30]. In this context, the dynamical problem is modeled by two evolution equations with respect to the free-surface elevation $\eta(\boldsymbol{x},t)$ and the trace of the potential on the free surface, $\psi(\boldsymbol{x},t)$, where $\boldsymbol{x}=(x_1,x_2)$ is the horizontal position. However, these two equations do not form a closed system, since they also contain the normal-to-the-free-surface fluid velocity. To close the system and march it in time, one needs to formulate and solve an appropriate Dirichlet-to-Neumann (DtN) operator [30], [31], [32]. Various perturbative methods have been developed to evaluate this DtN operator [31], [33], [34], [35], [36], [37], distinguished by the choice of the perturbed quantities and truncation procedures; their interrelation is discussed in [38]. The extension of these methods to variable bathymetry was first considered in [39] and [40], further developed in [41], [42], [43] and recently elaborated in [44]. The theory becomes quite complicated for varying bathymetry, while its perturbative character limits its applicability to domains which remains near to canonical ones.

Another line of work in the context of NLPF is to employ a convergent (non-perturbative) series expansion of the velocity potential in terms of prescribed vertical functions and unknown horizontal modal amplitudes, obtaining a dimensionally reduced reformulation of the hydrodynamic problem. This approach initiated in the case of linear water waves by Athanassoulis and Belibassakis [45], and extended later to fully nonlinear water waves by the same authors [46], [47], [48]. The wave potential is expanded, in any local vertical segment of the fluid domain, with respect to a standard $L^2-$basis augmented by two additional modes, the *sloping bottom* mode and the *free-surface* mode, that enable the consistent satisfaction of the bottom and the free-surface conditions. This expansion realizes an exact semi-separation of variables in the nonplanar, instantaneous fluid domain. A rigorous proof of the validity of this semi-separation of variables and its fast convergence, including the proof that the augmented vertical system is a basis in an $H^2$ (Sobolev) space, is recently given in [49]. Using this expansion, in conjunction with Luke's variational principle, a system of equations for the unknown modal amplitudes is derived, called consistent Coupled-Mode System (CMS). A drawback of the above approach is the presence of infinite (yet convergent) series in the evolution equations, which deviates from the Hamiltonian structure of the problem, complicating their numerical implementation, and making them inappropriate for long-time nonlinear simulations.

Recently, Papoutsellis and Athanassoulis [50] significantly advanced the above approach, reinstating the Hamiltonian structure of the evolution equations, by summing up the infinite series terms appearing in the consistent CMS formulation. This made possible by elaborating further on the Euler-Lagrange equations obtained by Luke's variational principle, resulting in two nonlinear, nonlocal, evolution equations, for the free-surface elevation and free-surface potential, which now contain only a nonlocal coefficient field, the modal amplitude of the free-surface mode, denoted by $\varphi_{-2} = \varphi_{-2}(x,t)$. The latter is obtained by solving an instantaneous, parametrically dependent on time, linear CMS of horizontal differential equations with respect to all modal amplitudes of the velocity potential. The coefficients of this CMS vary in space and time through their dependence on the free surface elevation and bathymetry. Its solution reproduces the DtN operator through a simple expression, derived in [49]. The above described evolutionary system together with the substrate CMS is subsequently called Hamiltonian Coupled-Mode System (HCMS) or Hamiltonian Coupled-Mode Theory (HCMT), according to the context. In a sense, HCMT can be considered as a reformulation of Zakharov and Craig & Sulem (ZCS) Hamiltonian equations, with different representation of the DtN operator. It should be stressed that the only assumptions needed for the derivation of HCMS are typical smoothness conditions on the physical boundaries of the fluid domain. No additional assumptions concerning the slope or the deformation of the free-surface elevation and the seabed are invoked, which ensures that HCMS accounts for fully non-linear and dispersive waves over variable bathymetry. Wave models of similar structure can also be obtained by approximating the wave potential in the vertical direction by means of Chebyshev polynomials, in conjunction with a transformation of the fluid domain to a flat strip [51], [52], [53].

The essential novelty of HCMT is the new representation of the DtN operator by means of the free-surface modal field $\varphi_{-2}$. This feature turns to be very important because, as also proved in [49], the modal characterization of the DtN operator exhibits a striking superconvergence for general geometries, including very steep seabed and free-surface shapes. Thus, truncating the convergent (non-asymptotic) series expansion of the wave potential at a small number of modes in the numerical solution scheme, usually not more than eight, suffices to produce very accurate results for strongly nonlinear problems.

The goal of this paper is twofold; first, to present a highly accurate implementation of the substrate kinematic CMS and, second, to apply the accurately implemented HCMS to demanding, nonlinear, water-wave problems over varying bathymetry, illuminating its ability for accurate simulation. The first goal is accomplished by the analytic calculation of all, space-time varying coefficients of the CMS in terms of the free-surface elevation $\eta(x,t)$ and the bathymetry $h(x)$. The analytic expressions of these coefficients, which are defined as integrals of the vertical basis functions over the intervals $[-h(x),\eta(x,t)]$, ensure their accurate evaluation (at machine accuracy) with very low computational cost. The second goal of this paper is effectuated by applying the HCMS to the simulation of various cases of solitary wave transformation over varying bathymetry both in two- and three-dimensional problems. Apart from validating our computations against existing experimental results and other computations, emphasis is also put on the consideration of new solitary wave-bottom interactions, demonstrating the capabilities of the HCMT to be used for investigating new phenomena, in which the interplay of nonlinearity and dispersion produces complicated wave patterns, involving various different space and time scales.

The organization of the paper is as follows. In Section 2, the standard formulation of the **NLPF** problem is recalled and the new HCMT is described bringing out the issues that are addressed in the main body of the paper. Section 3 presents the analytic calculation of the derivatives of the vertical basis functions used in the representation of the velocity potential. These preparatory results are used in Section 4 for the systematic derivation of analytic formulae for the variable coefficients of the substrate CMS in terms of $\eta(x,t)$ and $h(x)$. In Section 5, a numerical scheme for the solution of HCMS is described and its nonlinear accuracy and convergence is investigated against the steady propagation of a solitary wave over flat bottom. Section 6 is devoted to the study of several water-wave problems related to the transformation of solitary waves interacting with vertical walls and varying bathymetry. The first three examples refer to already studied cases. For these cases, our simulations reproduce with high accuracy various existing experimental measurements and computations by other authors, providing also some new results. The fourth example elaborates on a new type of "Bragg scattering" effect, appearing when a solitary wave propagates over a seabed with a sinusoidal patch; the fifth (last) example studies the disintegration and focusing of a solitary wave moving over a three-dimensional bathymetry consisting of parallel banks and troughs. In Section 7 a general discussion and the main conclusions are presented. Various technical issues, including an outline of the derivation of the HCMT, are discussed in five appendices. Finally, we mention that three videos are included, as electronic supplementary material, showing the evolution of the free-surface elevation for the last two examples. These videos can be viewed/downloaded by clicking the word video, appearing in subsections 6.4 and 6.5.

## 2. Formulation of the problem

### 2.1 Classical formulation of the problem

Consider a Cartesian coordinate system $Ox_1x_2z$, with $x_1x_2-$ plane coinciding with the quiescent free surface, and the $z$-axis pointing upwards. An ideal, homogenous and incompressible fluid fills the time-dependent domain

$$D_h^\eta = D_h^\eta(X,t) = \{(\boldsymbol{x},z) \in X \times \mathbb{R},\ -h(\boldsymbol{x}) < z < \eta(\boldsymbol{x},t)\},\quad t \in [t_0,t_1],$$

where $X \subset \mathbb{R}^d$ ($d = 1$ or $2$) is the common projection of the surfaces $z = \eta(\boldsymbol{x},t)$ and $z = -h(\boldsymbol{x})$ on the $x_1x_2-$ plane. $X$ is a fixed domain, since all finite lateral boundaries are assumed to be vertical. In the horizontal extent, the fluid domain $D_h^\eta(X,t)$ may be unbounded, partially bounded, or closed. For definiteness, we assume that the lateral boundary of $D_h^\eta(X,t)$ consists of three parts; a part $W_0 = \{(\boldsymbol{x},z): \boldsymbol{x} \in \partial X_0\}$ that is considered as a vertical impermeable wall; a part $W_e = \{(\boldsymbol{x},z): \boldsymbol{x} \in \partial X_e\}$ considered as an entrance boundary (a given wave system enters the computational domain through it); and what remains, denoted by $W_\infty = \{(\boldsymbol{x},z): \boldsymbol{x} \in \partial X_\infty\}$, extents up to infinity (radiation boundary). $\partial X_0, \partial X_e$ are either points (in the two-dimensional case), or lines on the $x_1x_2-$ plane (in the three-dimensional case). Any one (or two) of $W_0, W_e, W_\infty$ may be empty. Under the additional assumption of irrotationality, the fluid velocity in $D_h^\eta$ is described by a potential $\Phi = \Phi(\boldsymbol{x},z,t)$, also called wave potential, satisfying the following set of equations [54], Ch. 1:

$$\Delta\Phi = \left(\partial_{x_1}^2 + \partial_{x_2}^2 + \partial_z^2\right)\Phi = 0, \quad \text{in } D_h^\eta(X,t), \tag{1a}$$

$$\partial_t \eta + \nabla_{\boldsymbol{x}}\eta \cdot \left[\nabla_{\boldsymbol{x}}\Phi\right]_{z=\eta} - \left[\partial_z \Phi\right]_{z=\eta} = 0, \tag{1b}$$

$$\left[\partial_t \Phi\right]_{z=\eta} + \frac{1}{2}[\nabla\Phi]_{z=\eta}^2 + g\eta = 0, \tag{1c}$$

$$\nabla_{\boldsymbol{x}} h \cdot \left[\nabla_{\boldsymbol{x}}\Phi\right]_{z=-h} + \left[\partial_z \Phi\right]_{z=-h} = 0, \tag{1d}$$

supplemented by appropriate lateral boundary conditions, as follows:

$$n_{\partial X_0} \cdot \left[\nabla_x \Phi\right]_{(x,z) \in W_0} = 0, \tag{2a}$$

$$\begin{cases} \left[\eta(x,t)\right]_{x \in \partial X_e} \equiv \eta_e(x,t) \\ n_{\partial X_e} \cdot \left[\nabla_x \Phi(x,z,t)\right]_{(x,z) \in W_e} \equiv V_e(x,z,t) \\ \text{or} \quad \left[\Phi(x,z,t)\right]_{(x,z) \in W_e} \equiv \Phi_e(x,z,t) \end{cases}, \tag{2b}$$

$$\begin{cases} \left|\eta(x,t)\right|_{x \in \partial X_\infty} \to 0 \\ \left|\nabla_x \Phi(x,z,t)\right|_{(x,z) \in W_\infty} \to 0 \end{cases}, \quad \text{as} \quad \left|(x,z)\right|_{(x,z) \in W_\infty} \to \infty. \tag{2c}$$

In the above equations, $g$ is the acceleration of gravity, $\nabla_x = (\partial_{x_1}, \partial_{x_2})$ denotes the horizontal gradient, and $n_{\partial X_0}$, $n_{\partial X_e}$ denote the (horizontal) outward unit normal vector on $\partial X_0$ and $\partial X_e$, respectively. Boundary values (traces) are denoted by using brackets with a subscript specifying the boundary; for example, $\left[\partial_z \Phi\right]_{z=-h} = \partial_z \Phi(x, z=-h(x), t)$. Eq. (2b) defines the boundary conditions on the entrance boundary $W_e$. In this boundary, a prescribed free-surface elevation $\eta_e$ is required, supplemented by either the prescribed normal fluid velocity $V_e(x,z,t)$ or, alternatively, the prescribed values of the wave potential $\Phi_e(x,z,t)$, as indicated in Eq. (2b). The local depth of the fluid is

$$H(x,t) \equiv \eta(x,t) + h(x), \tag{3}$$

and is assumed to be positive at every $(x,t)$.

We mention here that the nonlinear problem (1) admits of an unconstrained variational formulation, introduced by Luke in 1967 [55]; see also [56], Ch. 13. The action functional of Luke's Variational Principle has the form

$$\mathcal{S}[\eta, \Phi] = \int_{t_0}^{t_1} \int_X L(\eta, \Phi) \, dx \, dt, \tag{4a}$$

where

$$L(\eta, \Phi) = \int_{-h}^{\eta} \left(\partial_t \Phi + \frac{1}{2} |\nabla_x \Phi|^2 + \frac{1}{2} (\partial_z \Phi)^2 + g z\right) dz + \\ + \begin{cases} \int_{-h}^{\eta_e} V_e(x,z,t) \left[\Phi\right]_{(x,z) \in W_e} dz, \\ 0, \end{cases} \tag{4b}$$

where the two alternatives appearing in the last term of the above equation correspond to the two alternative boundary conditions on the entrance boundary $W_e$, as stated in Eq. (2b). According to Luke's Variational Principle, the fields $\eta$ and $\Phi$ satisfy Eqs (1) if and only if they render the action functional $\mathcal{S}[\eta, \Phi]$ stationary, that is, they satisfy the variational equation

$$\delta \mathcal{S}[\eta,\Phi;\delta\Phi,\delta\eta] \;=\; \delta_\Phi \mathcal{S}[\eta,\Phi;\delta\Phi] + \delta_\eta \mathcal{S}[\eta,\Phi;\delta\eta] \;=\; 0. \tag{5}$$

The variational equation (5) provides a concise equivalent reformulation of the NLPF problem, Eqs. (1), (2). By using various approximate representations of the wave potential $\Phi$ in terms of series of known functions multiplied by unknown amplitudes, Eq. (5) generates approximate model equations for the NLPF problem. For example, this approach was used in [57] for the derivation of Boussinesq equations, and in [58] and [59] for the derivation of more general systems, including Boussinesq equations as a special case. In the present HCMT, Eq. (5) is used in conjunction with an exact, rapidly convergent, series expansion of the wave potential resulting in the HCMS, as explained in the next section.

**2.2. The Hamiltonian Coupled-Mode Theory**

**2.2.1** *Exact vertical series expansion of the wave potential*

As established by the present authors in [49], the wave potential can be represented in the form of the following series expansion, realizing an *exact semi-separation of variables* in the irregular, instantaneous fluid domain $D_h^\eta(X,t)$:

$$\Phi(\boldsymbol{x},z,t) = \varphi_{-2}(\boldsymbol{x},t)Z_{-2}(z;\eta,h) + \varphi_{-1}(\boldsymbol{x},t)Z_{-1}(z;\eta,h) \\ + \sum_{n=0}^{\infty} \varphi_n(\boldsymbol{x},t)Z_n(z;\eta,h), \tag{6}$$

where $Z_n$ are explicitly given functions in terms of the local values of the free-surface elevation $\eta(\boldsymbol{x},t)$ and bathymetry $h(\boldsymbol{x})$ (see below), and $\varphi_n$ are unknown modal amplitudes. Under plausible smoothness assumptions on the boundary functions $\eta(\boldsymbol{x},t)$ and $h(\boldsymbol{x})$, the series in the right-hand side of Eq. (6) converges absolutely and rapidly to the wave potential $\Phi$, and the series obtained by term-wise differentiation converge (absolutely and rapidly as well) to the corresponding derivatives of $\Phi$, throughout the whole fluid domain $D_h^\eta(X,t)$, up to and including the boundaries. For rigorous proofs of the above statements see [49], Theorems 1 and 2, and corresponding Corollaries.

The *vertical functions* $Z_n(z;\eta,h)$, $n \geq -2$, are defined so that to satisfy the normalization condition

$$\left[Z_n\right]_{z=\eta} = 1, \quad n \geq -2, \tag{7}$$

which implies

$$\sum_{n=-2}^{\infty} \varphi_n = \left[\Phi\right]_{z=\eta}. \tag{8}$$

Their explicit expressions are as follows (see [49]):

$$Z_n(z;\eta,h) = \alpha_n^{(2)}(z+h)^2 + \alpha_n^{(1)}(z+h) + a_n^{(0)}, \qquad n=-2,-1, \tag{9a}$$

$$Z_0(z;\eta,h) = \frac{\cosh[k_0(z+h)]}{\cosh(k_0 H)}, \quad Z_n(z;h,\eta) = \frac{\cos[k_n(z+h)]}{\cos(k_n H)}, \quad n \geq 1, \tag{9b,c}$$

where the quantities $k_n = k_n(\eta, h) = k_n(x, t)$, $n \geq 0$, are positive functions implicitly dependent on $(\eta, h)$ through the transcendental equations

$$\mu_0 - k_0 \tanh(k_0 H) = 0, \qquad \mu_0 + k_n \tan(k_n H) = 0, \, n \geq 1, \qquad (10a,b)$$

and $H = H(x, t) = \eta(x, t) + h(x)$ is the local depth at $(x, t)$. Eqs. (10a,b) will be called in the sequel *local dispersion relations*, and their solutions $k_n$ will be referred to as *local wavenumbers* ([1]).

The coefficients defining the functions $Z_{-2}$, $Z_{-1}$, Eq. (9a), are given by

$$\alpha_{-2}^{(2)} = \frac{a_{-2}}{H(x,t)}, \qquad \alpha_{-2}^{(1)} = 0, \qquad \alpha_{-2}^{(0)} = -a_{-2} H(x,t) + 1, \qquad (11a)$$

$$\alpha_{-1}^{(2)} = \frac{a_{-1}}{H(x,t)}, \qquad \alpha_{-1}^{(1)} = \frac{1}{h_0}, \qquad \alpha_{-1}^{(0)} = a_{-2} H(x,t) - 1, \qquad (11b)$$

where

$$a_{-2} = (\mu_0 h_0 + 1)/2 h_0, \qquad a_{-1} = (\mu_0 h_0 - 1)/2 h_0. \qquad (11c)$$

The quantities $\mu_0$ and $h_0$, appearing in Eqs. (10a,b) and (11c), are auxiliary positive constants, whose values do not affect the validity of the HCMT, although they may affect its numerical behavior. Suggestions concerning their efficient choice are given below, in Remarks 2.2 and 2.3.

**Remarks: 2.1.** The functions $Z_n(z; \eta, h)$, $n \geq 0$, are defined as the eigenfunctions of the $(x, t)$-local vertical Sturm-Liouville problem

$$\partial_z^2 Z_n(z) + k_n^2 Z_n(z) = 0, \qquad -h(x) < z < \eta(x, t), \qquad (12a)$$

with boundary conditions

$$[(\partial_z - \mu_0) Z_n]_{z=\eta} = 0, \qquad \text{and} \qquad [\partial_z Z_n]_{z=-h} = 0, \qquad (12b)$$

under the normalization (7). The set of these eigenfunctions, $\{Z_n, n \geq 0\}$, forms a basis in the space $L^2(-h(x), \eta(x, t))$, which makes tempting to consider the following series expansion of the wave potential $\Phi(x, z, t)$ at each vertical segment $[-h(x), \eta(x, t)]$ of the fluid domain:

$$\Phi(x, z, t) = \sum_{n=0}^{\infty} \varphi_n(x, t) Z_n(z; \eta, h). \qquad (13)$$

Such an expansion converges only in $L^2$ – sense, thus very slowly, and does not provide the correct values of $\partial_z \Phi$ at the end points $z = -h(x)$ and $z = \eta(x, t)$, except if the potential itself satisfies conditions (12b). This deficiency of the expansion (13) has been discussed by many authors; see e.g. [45], [48] [60], [61], [49], and references cited therein. To ensure an exact representation of $\Phi$ and its derivatives, and accelerate the convergence rate for arbitrary (smooth) shape of the surfaces $z = -h(\mathbf{x})$ and $z = \eta(\mathbf{x},t)$, we have to add to the right-hand side of the expansion (13) two additional "boundary modes" $\varphi_{-2} Z_{-2}$ and $\varphi_{-1} Z_{-1}$.

---

([1]) The terminology comes from the linear water-wave theory, although there is not any linearization herein.

**2.2.** The vertical functions $Z_{-2}$ and $Z_{-1}$ are chosen in such a way so that to remove the restrictions (12b), imposed by the properties of the eigenfunctions $Z_n$, $n \geq 0$. Thus, they have to satisfy the conditions $[(\partial_z - \mu_0) Z_{-2}]_{z=\eta} \neq 0$ and $[\partial_z Z_{-1}]_{z=-h} \neq 0$. For definiteness, they are chosen as the least degree polynomials satisfying

$$[(\partial_z - \mu_0) Z_{-1}]_{z=\eta} = 0, \qquad [\partial_z Z_{-1}]_{z=-h} = 1/h_0, \qquad (14a,b)$$

$$[(\partial_z - \mu_0) Z_{-2}]_{z=\eta} = 1/h_0, \qquad [\partial_z Z_{-2}]_{z=-h} = 0, \qquad (14c,d)$$

and the normalization condition (7). (The constant $h_0$ is introduced for dimensional purposes). In this way, we get the forms given by Eq. (9a). Other choices of the vertical functions $Z_{-2}$ and $Z_{-1}$, satisfying the same conditions (14), produce equivalent systems, leading to the same results. This is also theoretically expected since, as proved in [49], the system of functions $\left\{ Z_{-2}, Z_{-1}, Z_n/k_n^2, n \geq 0 \right\}$ forms a Riesz basis in the Sobolev space $H^2\left(-h(\mathbf{x}), \eta(\mathbf{x},t)\right)$, at each $(\mathbf{x},t)$.

**2.3.** Functions $Z_n(z;\eta,h)$ depend on the numerical parameter $h_0$ and $\mu_0$. $h_0$, introduced only for dimensional purposes, may be taken as a characteristic depth of the studied configuration, e.g. the depth at the incident region, or the mean depth. The constant $\mu_0$ is introduced in the formulation of the Sturm-Liouville problem, defining the eigenfunctions $\{Z_n\}_{n \geq 0}$, Eqs. (12b). All theoretical statements made above remain valid for any value $\mu_0 > 0$. Since, however, for a specific choice of $\mu_0$ the eigenfunctions $\{Z_n\}_{n \geq 0}$ become the physical modes of linear waves with angular frequency $\omega_0 = \sqrt{g \mu_0}$ (at the local depth), it is preferable to select $\mu_0$ in relation with a characteristic frequency of the problem, e.g. the frequency of the incident wave, or the central frequency of a wave packet or wave spectrum. In this way, series expansion (6) encapsulates the physics of a "nearby" linear wave problem a priori, that is, before using the dynamical evolution equations (see Eqs. (16)), by means of which the modal amplitudes $\varphi_n(\mathbf{x},t)$ will be determined taking fully into account all nonlinear features of the problem. Such a choice of $\mu_0$ improves the accuracy of the calculations, for a given number of modes in the truncated series used in the numerical simulations.

### 2.2.2 *The Hamiltonian coupled-mode system*

Representation (6) establishes the change of functional variables

$$\left(\eta(\mathbf{x},t), \Phi(\mathbf{x},z,t)\right) \leftrightarrow \left(\eta(\mathbf{x},t), \{\varphi_n(\mathbf{x},t)\}_{n=-2}^{\infty}\right), \qquad (15)$$

which can be employed for the dimensional reduction (elimination of the $z$ variable) of the nonlinear free surface problem (1). Introducing the representation (6) of the wave potential into the variational equation (5), and performing the variations with respect to the new independent functional variables $\eta(\mathbf{x},t)$ and $\varphi_n(\mathbf{x},t)$, $n \geq -2$, we eventually obtain, after an extensive analytical treatment presented in detail in [50], the following two Hamiltonian evolution equations with respect to $\eta(x,t)$ and $\psi(x,t) = [\Phi]_{z=\eta} = \sum_{n=-2}^{\infty} \varphi_n$,

$$\partial_t \eta = -(\nabla_x \eta) \cdot (\nabla_x \psi) + (|\nabla_x \eta|^2 + 1)(h_0^{-1} \varphi_{-2} + \mu_0 \psi), \tag{16a}$$

$$\partial_t \psi = -g\eta - \frac{1}{2}(\nabla_x \psi)^2 + \frac{1}{2}(|\nabla_x \eta|^2 + 1)(h_0^{-1} \varphi_{-2} + \mu_0 \psi)^2, \tag{16b}$$

and the following system of horizontal differential equations with respect to the modal amplitude $\varphi_n$, $n \geq -2$, at each $(x,t)$:

$$\sum_{n=-2}^{\infty} \left( A_{m,n} \nabla_x^2 + \boldsymbol{B}_{m,n} \cdot \nabla_x + C_{m,n} \right) \varphi_n = 0, \qquad x \in X, \quad m \geq -2, \tag{17a}$$

$$\sum_{n=-2}^{\infty} \varphi_n = \psi, \qquad x \in X, \tag{17b}$$

supplemented by appropriate lateral boundary conditions on $\partial X$. The coefficients $A_{m,n}$, $\boldsymbol{B}_{m,n} = (B_{m,n}^1, B_{m,n}^2)$ and $C_{m,n}$, appearing in the left-hand side of Eq. (17a), are given by the formulae

$$A_{m,n} = \int_{-h}^{\eta} Z_n Z_m \, dz, \tag{18a}$$

$$\boldsymbol{B}_{m,n} = 2 \int_{-h}^{\eta} \nabla_x Z_n Z_m \, dz + \nabla_x h \left[ Z_n Z_m \right]_{z=-h}, \tag{18b}$$

$$C_{m,n} = \int_{-h}^{\eta} \left( \Delta_x Z_n + \partial_z^2 Z_n \right) Z_m \, dz + \left( \nabla_x h, 1 \right) \cdot \left[ \begin{pmatrix} \nabla_x Z_n \\ \partial_z Z_n \end{pmatrix} Z_m \right]_{z=-h}. \tag{18c}$$

A sketch of the derivation of Eqs. (16) and (17), along with the modal form of the lateral boundary conditions on $\partial X$, Eqs. (2), is presented in Appendix A. The following remarks are in order here.

**Remarks: 2.4.** The evolution equations (16) are not closed with respect to $\eta(x,t)$ and $\psi(x,t)$ since they contain the free-surface modal amplitude $\varphi_{-2}$. The latter is provided by solving the system of equations (17a,b) at each $(x,t)$. Since the coefficients of this system are defined in terms of $\eta(x,t)$ and $h(x)$ (see Eqs. (18)), and its excitation is $\psi$ (see Eq. (17b)), it is expedient to write

$$\varphi_{-2} = \mathcal{F}[\eta, h]\psi, \tag{19}$$

revealing that $\varphi_{-2}$ is in fact a linear, nonlocal operator on $\psi$, also dependent (nonlinearly) on the boundary functions $\eta$ and $h$.

**2.5.** The operator $\mathcal{F}[\eta, h]\psi$ is similar in nature to the classical DtN operator $\mathcal{G}[\eta, h]\psi$, as defined in [31], their Eq. (2.4). In fact, as shown in [49], the following identity holds true,

$$\mathcal{G}[\eta,h]\psi = -\nabla_x\eta\cdot\nabla_x\psi + \left(\left|\nabla_x\eta\right|^2 + 1\right)\left(\mathcal{F}[\eta,h]\psi/h_0 + \mu_0\psi\right)^2. \tag{20}$$

This identity permits us to show that the evolution Eqs. (16) are equivalent to the classical ZCS formulation, as presented by Eqs. (2.5) of [31].

**2.6.** Having solved the evolution equations (16), the calculation of any field quantities (pressure, velocity, acceleration) throughout the whole domain is almost *costless*, since it can be performed by simple manipulation of the rapidly convergent series (6).

A key issue for the efficient implementation of the HCMT, Eqs. (16) and (17), is the accurate and fast calculation of the coefficients $A_{m,n}, B_{m,n}, C_{m,n}$, which are evaluated at each $(x,t)$. As seen by Eqs. (18), these coefficients are expressed as vertical integrals of $Z_n$ multiplied by some derivatives of $Z_n$. Thus, the best way towards an efficient calculation of the coefficients is the explicit analytic calculation of the involved derivatives and integrals. The relevant results are systematically presented in Sec. 3 and 4. These results permit us to express analytically all coefficients $A_{m,n}$, $B_{m,n}$, $C_{m,n}$ in terms of $\eta(x,t)$, $h(x)$, $h_0$, $\mu_0$ and the roots $k_n$ of transcendental equations (10a,b). The latter are evaluated by a semi-analytic iterative method, ensuring machine accuracy with no more than 3 iterations, as proved in [62]. Accordingly, all coefficients are very accurately calculated without using any numerical procedure concerning differentiation, integration or root finding.

## 3. Calculation of the spatial derivatives of the vertical basis functions

The purpose of this section is to derive exact, closed-form expressions for the first- and second-order spatial derivatives of $Z_n = Z_n(z;\eta(x,t),h(x))$, Eqs. (9), in terms of $Z_n$ themselves and the roots $k_n$ of the transcendental equations (10)([2]). These expressions will be presented in a form facilitating their exploitation, in the next section, for the analytic calculation of the vertical integrals appearing in the coefficients $A_{m,n}$, $B_{m,n}$, $C_{m,n}$, Eqs. (18).

It should be stressed herein that no simplifying assumptions are invoked in all analytic calculations of this and the next section. The results presented are exact for any shape of the free surface elevation $\eta(x,t)$ and bathymetry $h(x)$. That being said, it is clear that the present calculations embody and generalize various similar results already obtained in the context of the implementation of various mild-slope and coupled-mode equations for water waves; see e.g. [63], [45], [59].

### 3.1 Spatial derivatives of $Z_n$, $n = -2, -1$

The horizontal gradients of the functions $Z_{-2}$ and $Z_{-1}$, Eqs. (9a), are easy to calculate. They are written here in the form

$$\nabla_x Z_n = \boldsymbol{F}_n^{(1)} + \boldsymbol{F}_n^{(2)}(z+h) + \boldsymbol{F}_n^{(3)}(z+h)^2, \quad n = -2,-1, \tag{21}$$

---

([2]) This statement does not apply to the derivatives of $Z_{-2}$, $Z_{-1}$. These vertical functions are polynomials in $z$, and their derivatives are calculated in a straightforward way.

where $\boldsymbol{F}_n^{(k)} \equiv \boldsymbol{F}_n^{(k)}(\boldsymbol{x},t)$, $k = 1, 2, 3$, are horizontal vector fields given by

$$\boldsymbol{F}_n^{(1)} = \nabla_{\boldsymbol{x}} \alpha_n^{(0)} + \alpha_n^{(1)} \nabla_{\boldsymbol{x}} h, \qquad \boldsymbol{F}_n^{(2)} = 2\alpha_n^{(2)} \nabla_{\boldsymbol{x}} h, \qquad \boldsymbol{F}_n^{(3)} = \nabla_{\boldsymbol{x}} \alpha_n^{(2)}, \qquad (22)$$

and $\alpha_n^{(0)}, \alpha_n^{(1)}, \alpha_n^{(2)}$ are defined by Eqs. (11). The form of presentation of $\nabla_{\boldsymbol{x}} Z_n$, Eq. (21), as a polynomial with respect to $z+h$, is dictated by its use, in the next section, for calculating the vertical integrals appearing in Eqs. (18).

Taking the horizontal divergence of both members of Eq. (21), we obtain the horizontal Laplacians of $Z_n$, $n = -2, -1$, in the form

$$\Delta_{\boldsymbol{x}} Z_n = G_n^{(1)} + G_n^{(2)}(z+h) + G_n^{(3)}(z+h)^2, \qquad n = -2, -1, \qquad (23)$$

where the scalar fields $G_n^{(k)} \equiv G_n^{(k)}(\boldsymbol{x},t)$ are given by

$$G_n^{(1)} = \nabla_{\boldsymbol{x}} \cdot \boldsymbol{F}_n^{(1)} + \boldsymbol{F}_n^{(2)} \cdot \nabla_{\boldsymbol{x}} h, \qquad (24a)$$

$$G_n^{(2)} = \nabla_{\boldsymbol{x}} \cdot \boldsymbol{F}_n^{(2)} + 2\boldsymbol{F}_n^{(3)} \cdot \nabla_{\boldsymbol{x}} h, \qquad G_n^{(3)} = \nabla_{\boldsymbol{x}} \cdot \boldsymbol{F}_n^{(3)}. \qquad (24b,c)$$

Finally, the first and second vertical derivatives of $Z_{-2}$ and $Z_{-1}$ are

$$\partial_z Z_{-1} = 2\alpha_{-1}^{(2)}(z+h) + \alpha_{-1}^{(1)}, \qquad \partial_z Z_{-2} = 2\alpha_{-2}^{(2)}(z+h), \qquad (25a)$$

$$\partial_z^2 Z_n = 2\alpha_n^{(2)}, \qquad n = -2, -1. \qquad (25b)$$

### 3.2 Spatial derivatives of $Z_n$, $n \geq 0$

For the calculation of the horizontal derivatives of $Z_n$, $n \geq 0$, we have to take into account that the functions $Z_n$ are dependent on $\boldsymbol{x}$ only through their dependence on the fields $\eta = \eta(\boldsymbol{x},t)$ and $h = h(\boldsymbol{x})$. Thus, by using the chain rule, we obtain the relation

$$\nabla_{\boldsymbol{x}} Z_n = \partial_\eta Z_n \nabla_{\boldsymbol{x}} \eta + \partial_h Z_n \nabla_{\boldsymbol{x}} h, \qquad n \geq 0, \qquad (26)$$

where $\partial_\eta Z_n$ and $\partial_h Z_n$ are the derivatives of the eigenfunctions $Z_n$ with respect to the local values of the free-surface elevation $\eta$ and the depth $h$, respectively, at each $(\boldsymbol{x},t)$. In calculating the derivatives $\partial_\eta Z_n$ and $\partial_h Z_n$, we have to take into account both the explicit dependence of $Z_n$ on $(\eta, h)$ and the implicit dependence through the quantity $k_n = k_n(\eta, h)$, involved in the definition of $Z_n$; see Eqs. (9b,c). Thus, we also need to calculate the derivatives $\partial_\eta k_n$ and $\partial_h k_n$, by means of Eqs. (10), which define implicitly the functions $k_n = k_n(\eta, h)$. For $n \geq 1$ these implicit differentiations have been performed in [49], Sec. 3, Proposition 3.1. The case $n = 0$ is treated similarly. The final results, for all $n \geq 0$, read as follows:

$$\partial_\eta k_n = \partial_h k_n = \partial_H k_n = \begin{cases} -\dfrac{k_0(k_0^2 - \mu_0^2)}{\mu_0 + H(k_0^2 - \mu_0^2)}, & n = 0, \\[2mm] \dfrac{k_n(k_n^2 + \mu_0^2)}{\mu_0 - H(k_n^2 + \mu_0^2)}, & n \geq 1, \end{cases} \tag{27a}$$

$$\partial_\eta^2 k_n = \partial_h^2 k_n = \partial_{\eta h} k_n = \partial_H^2 k_n =$$
$$= -2\partial_H k_n \left\{ \mu_0 + \frac{\partial_H k_n}{k_n}(H\mu_0 - 1)\left[2 + H\frac{\partial_H k_n}{k_n}\right]\right\}, \quad n \geq 0. \tag{27b}$$

where $H = \eta + h$; see Eq. (3). From the above equations we see that all derivatives of $k_n$ with respect to $\eta$ and $h$ are expressed in terms of $k_n$ itself and $H$. Use will be also made, in the sequel, of the following equations, which are obtained in a straightforward way:

$$\nabla_x k_n = \partial_H k_n \nabla_x H \quad \text{and} \quad \Delta_x k_n = \partial_H^2 k_n (\nabla_x H)^2 + \partial_H k_n \Delta_x H. \tag{28}$$

Before proceeding with the calculation of $\nabla_x Z_n$, it is convenient to introduce the auxiliary functions

$$W_0 = \frac{\sinh[k_0(z+h)]}{\cosh(k_0 H)}, \qquad W_n = \frac{\sin[k_n(z+h)]}{\cos(k_n H)}, \qquad n \geq 1, \tag{29a,b}$$

- **The case $n = 0$**

The partial derivatives $\partial_\eta Z_0$ and $\partial_h Z_0$ are calculated by differentiating Eq. (9b) with respect to $\eta$, $h$, taking into account the implicit dependence $k_0 = k_0(\eta, h)$, as explained above. The use of the local dispersion relation, Eq. (10a), permits as to eliminate the hyperbolic function $\tanh(k_0 H)$, as shown below:

$$\partial_\eta Z_0 = -\tanh(k_0 H)(\partial_H k_0 H + k_0)Z_0 + \partial_H k_0(z+h)W_0 =$$
$$= -\left(1 + H k_0^{-1}\partial_H k_0\right)\mu_0 Z_0 + \partial_H k_0(z+h)W_0, \tag{30a}$$

$$\partial_h Z_0 = -\tanh(k_0 H)(\partial_H k_0 H + k_0)Z_0 + k_0 W_0 + \partial_H k_0(z+h)W_0$$
$$= -\left(1 + \frac{H\partial_H k_0}{k_0}\right)\mu_0 Z_0 + k_0 W_0 + \partial_H k_0(z+h)W_0. \tag{30b}$$

Combining Eqs. (26), (27) and (30), we obtain

$$\nabla_x Z_0 = \boldsymbol{F}_0^{(1)} Z_0 + \boldsymbol{F}_0^{(2)} W_0 + \boldsymbol{F}_0^{(3)}(z+h) W_0, \tag{31}$$

where $\boldsymbol{F}_0^{(k)} = \boldsymbol{F}_0^{(k)}(\boldsymbol{x}, t)$, $k = 1, 2, 3$, are given by

$$\boldsymbol{F}_0^{(1)} = -\mu_0 \frac{\nabla_x(k_0 H)}{k_0}, \tag{32a}$$

$$\boldsymbol{F}_0^{(2)} = k_0 \nabla_x h, \qquad \boldsymbol{F}_0^{(3)} = \nabla_x k_0. \tag{32b,c}$$

To calculate $\Delta_x Z_0$ we can apply the $\nabla_x$ operator to both members of Eq. (31). Thus, the gradient of the function $W_0$ is required. Working similarly as above, we find

$$\nabla_x W_0 = F_0^{(1)} W_0 + F_0^{(2)} Z_0 + F_0^{(3)} (z+h) Z_0, \tag{33}$$

where the fields $F_0^{(1)}$, $F_0^{(2)}$, $F_0^{(3)}$ are again given by Eqs. (32). Taking now the horizontal gradient of both members of Eq. (31), we obtain

$$\Delta_x Z_0 = \left(\nabla_x \cdot F_0^{(1)}\right) Z_0 + F_0^{(1)} \cdot \left(\nabla_x Z_0\right) + \left(\nabla_x \cdot F_0^{(2)}\right) W_0 + F_0^{(2)} \cdot \left(\nabla_x W_0\right)$$
$$+ \left(\nabla_x \cdot F_0^{(3)}\right)(z+h) W_0 + F_0^{(3)} \cdot \left(\nabla_x h\right) W_0 + F_0^{(3)} \cdot \left(\nabla_x W_0\right)(z+h)$$

which, after substitution of $\nabla_x Z_0$, $\nabla_x W_0$ from Eqs. (31) and (33) and straightforward manipulations, yields

$$\Delta_x Z_0 = G_0^{(1)} Z_0 + G_0^{(2)} (z+h) Z_0 + G_0^{(3)} (z+h)^2 Z_0 + G_0^{(4)} W_0 + G_0^{(5)} (z+h) W_0, \tag{34}$$

where

$$G_0^{(1)} = \left|F_0^{(1)}\right|^2 + \left|F_0^{(2)}\right|^2 + \nabla_x \cdot F_0^{(1)}, \tag{35a}$$

$$G_0^{(2)} = 2 F_0^{(2)} \cdot F_0^{(3)}, \tag{35b}$$

$$G_0^{(3)} = \left|F_0^{(3)}\right|^2, \tag{35c}$$

$$G_0^{(4)} = 2 F_0^{(2)} \cdot F_0^{(1)} + \nabla_x \cdot F_0^{(2)} + F_0^{(3)} \cdot (\nabla_x h), \tag{35d}$$

$$G_0^{(5)} = 2 F_0^{(3)} \cdot F_0^{(1)} + \nabla_x \cdot F_0^{(3)}. \tag{35e}$$

The first and second vertical derivatives of $Z_0$ are given by

$$\partial_z Z_0 = k_0 W_0, \qquad \partial_z^2 Z_0 = k_0^2 Z_0, \tag{36a,b}$$

**- The case $n \geq 1$**

Analytic expressions for the spatial derivatives of $Z_n$, $n \geq 1$, Eq. (9c), have been derived in [49]. In conformity with the present notation, the horizontal gradient of $Z_n$, $n \geq 1$ is written as

$$\nabla_x Z_n = F_n^{(1)} Z_n + F_n^{(2)} W_n + F_n^{(3)} (z+h) W_n, \qquad n \geq 1, \tag{37}$$

where

$$F_n^{(1)} = F_n^{(1)}(x,t) = -\mu_0 \frac{\nabla_x (k_n H)}{k_n}, \tag{38a}$$

$$F_n^{(2)} = F_n^{(2)}(x,t) = -k_n \nabla_x h, \qquad F_n^{(3)} = F_n^{(3)}(x,t) = -\nabla_x k_n. \tag{38b,c}$$

Again, for the calculation of the horizontal Laplacian, the horizontal gradient of the functions $W_n$, $n \geq 1$, Eq. (29b), is required. Working similarly as above, we find

$$\nabla_x W_n = F_n^{(1)} W_n - F_n^{(2)} Z_n - F_n^{(3)} (z+h) Z_n, \qquad n \geq 1, \tag{39}$$

where the fields $F_n^{(1)}$, $F_n^{(2)}$, $F_n^{(3)}$ are given by Eqs. (38). Taking the horizontal gradient of both members of Eq. (37), we obtain

$$\Delta_x Z_n = G_n^{(1)} Z_n + G_n^{(2)}(z+h) Z_n + G_n^{(3)}(z+h)^2 Z_n + G_n^{(4)} W_n + G_n^{(5)}(z+h) W_n, \quad n \geq 1 \tag{40}$$

where

$$G_n^{(1)} = \left|F_n^{(1)}\right|^2 - \left|F_n^{(2)}\right|^2 + \nabla_x \cdot F_n^{(1)}, \tag{41a}$$

$$G_n^{(2)} = -2 F_n^{(2)} \cdot F_n^{(3)}, \qquad G_n^{(3)} = -\left|F_n^{(3)}\right|^2 \tag{41b,c}$$

$$G_n^{(4)} = 2 F_n^{(1)} \cdot F_n^{(2)} + \nabla_x \cdot F_n^{(2)} + \nabla_x h \cdot F_n^{(3)}, \tag{41d}$$

$$G_n^{(5)} = 2 F_n^{(1)} \cdot F_n^{(3)} + \nabla_x \cdot F_n^{(3)}. \tag{41e}$$

The vertical derivatives are given by

$$\partial_z Z_n = -k_n W_n, \qquad \partial_z^2 Z_n = -k_n^2 Z_n, \qquad n \geq 1. \tag{42a,b}$$

### 3.3 Discussion of the above results

In this section, we have obtained explicit, closed-form expressions of the derivatives $\nabla_x Z_n$, $\Delta_x Z_n$, $\partial_z Z_n$, $\partial_z^2 Z_n$, appearing in the right-hand side of Eqs. (18), for all $Z_n$, $n \geq -2$. These closed-form formulae, permits us to avoid numerical differentiation, expressing the derivatives of $Z_n$ in terms of: (a) the vertical functions $Z_n$ themselves and the auxiliary functions $W_n$, (b) the boundary fields $\eta(x,t)$ and $h(x)$, and (c) the roots $k_n(x,t)$, $n \geq 0$ of the local dispersion relations, Eq. (10). These expressions are given in forms appropriate for application to the analytic calculations of vertical integrals, Eqs. (18). Using them, we are able to reduce the calculation of all vertical integrals to the calculation of the following basic ones

$$J(s;f) = \int_{z=-h(x)}^{z=\eta(x,t)} (z+h)^s f(z,t) dz = \int_{u=0}^{u=H} u^s f(u,t) du, \tag{43}$$

for $f(z,t) = Z_n Z_m, W_n Z_m, Z_n$, and $s = 0,1,2$, as shown in the next section.

### 4. Calculation of the coefficients of the substrate modal system

We shall now proceed with the analytic calculation of the $(x,t)-$ dependent coefficients $A_{m,n}$, $B_{m,n}$, $C_{m,n}$, $m,n \geq -2$, Eqs. (18). These coefficients are composed by integral and boundary terms (denoted, respectively, by the superscript "int" and "b"), written here as

$$A_{m,n} = A_{m,n}^{\text{int}}, \qquad B_{m,n} = B_{m,n}^{\text{int}} + B_{m,n}^{\text{b}}, \qquad C_{m,n} = C_{m,n}^{\text{int}} + C_{m,n}^{\text{b}}, \tag{44}$$

with

$$B_{m,n}^{\text{int}} = 2 \int_{-h}^{\eta} (\nabla_x Z_n) Z_m \, dz, \qquad B_{m,n}^{\text{b}} = (\nabla_x h)\left[Z_n Z_m\right]_{-h}, \tag{45a,b}$$

$$C^{\text{int}}_{m,n} = \int_{-h}^{\eta} (\nabla_x^2 Z_n + \partial_z^2 Z_n) Z_m \, dz, \qquad C^{\text{b}}_{m,n} = N_h \cdot \left[(\nabla_x Z_n, \partial_z Z_n) Z_m\right]_{-h}. \qquad (46\text{a,b})$$

In this section, closed-form analytic expressions for all these coefficients are derived, in terms of the local depth $H(x,t) = \eta(x,t) + h(x)$ and the local eigenvalues $k_n = k_n(x,t)$, the later defined implicitly by means of the local dispersion relations (10a,b). The formulae become concise and easy to implement numerically by using the notation (43), that is, by expressing the coefficients in terms of the basic integrals $J(s; Z_n)$, $J(s; Z_n Z_m)$ and $J(s; W_n Z_m)$. Analytic calculations of the latter integrals, in terms of $H$ and $k_n$, are systematically presented in Appendix B.

### 4.1 Reduction to the basic integrals

Taking into account the expressions for the horizontal gradient $\nabla_x Z_n$, Eqs. (21), (31), (37), the horizontal Laplacian $\Delta_x Z_n$, Eqs. (23), (34), (40), and the vertical derivatives $\partial_z Z_n$, $\partial_z^2 Z_n$, Eqs. (25), (36), (42), in conjunction with Eq. (43), we obtain

$$A_{m,n} = J(0; Z_n Z_m), \qquad n, m \geq -2, \tag{47}$$

$$B^{\text{int}}_{m,n} = 2\left(F_n^{(1)} J(0; Z_m) + F_n^{(2)} J(1; Z_m) + F_n^{(3)} J(2; Z_m)\right), \quad \begin{cases} n = -2, -1 \\ m \geq -2 \end{cases} \tag{48a}$$

$$B^{\text{int}}_{m,n} = 2\left(F_n^{(1)} J(0; Z_n Z_m) + F_n^{(2)} J(0; W_n Z_m) + F_n^{(3)} J(1; W_n Z_m)\right), \quad \begin{cases} n \geq 0 \\ m \geq -2 \end{cases} \tag{48b}$$

$$C^{\text{int}}_{m,n} = \left(G_n^{(1)} + \frac{2a_n}{H}\right) J(0; Z_m) + G_n^{(2)} J(1; Z_m) + G_n^{(3)} J(2; Z_m), \quad \begin{cases} n = -2, -1 \\ m \geq -2 \end{cases} \tag{49a}$$

$$C^{\text{int}}_{m,n} = \left(G_0^{(1)} + k_0^2\right) J(0; Z_0 Z_m) + G_0^{(2)} J(1; Z_0 Z_m) + \\ G_0^{(3)} J(2; Z_0 Z_m) + G_0^{(4)} J(0; W_0 Z_m) + G_0^{(5)} J(1; W_0 Z_m), \quad \begin{cases} n = 0 \\ m \geq -2 \end{cases} \tag{49b}$$

$$C^{\text{int}}_{m,n} = \left(G_n^{(1)} - k_n^2\right) J(0; Z_n Z_m) + G_n^{(2)} J(1; Z_n Z_m) + \\ + G_n^{(3)} J(2; Z_n Z_m) + G_n^{(4)} J(0; W_n Z_m) + G_n^{(5)} J(1; W_n Z_m), \quad \begin{cases} n \geq 1 \\ m \geq -2 \end{cases} \tag{49c}$$

$$B^{\text{b}}_{m,n} = \nabla_x h \left[Z_n Z_m\right]_{-h}, \qquad n, m \geq -2 \tag{50}$$

$$C^{\text{b}}_{m,n} = -\nabla_x h \cdot \left[\nabla_x Z_n Z_m\right]_{-h} - \left[\partial_z Z_n Z_m\right]_{-h} = \begin{cases} F_n^{(1)} [Z_m]_{-h}, & n = -2, -1, \ m \geq -2, \\ F_n^{(1)} [Z_n Z_m]_{-h}, & n \geq 0, \ m \geq -2. \end{cases} \tag{51}$$

## 4.2 Explicit formulae for some basic vertical integrals

For illustration purposes, explicit formulae are given in this subsection for the basic vertical integrals $J(0; Z_n Z_m)$, $m, n \geq 1$, and $J(s; W_n Z_m)$, $m, n \geq 1$, $s = 0, 1$. This sample of results illustrates the premise that the basic integrals, and thus the coefficients themselves, are only dependent on the local depth $H(\boldsymbol{x}, t)$ and local eigenvalues $k_n(\boldsymbol{x}, t)$. In addition, these formulae are quite simple, in contrast with some other ones, e.g. $J(2; Z_n Z_m)$, which become inconveniently large and they are better treated by using intermediate auxiliary formulae, as presented in Appendix B. For $J(0; Z_n Z_m)$ we have

$$J(0; Z_n Z_m) = \begin{cases} \dfrac{-\mu_0 + H(k_m^2 + \mu_0^2)}{2 k_m^2}, & m = n \geq 1, \\ 0, & m, n \geq 1, \ m \neq n. \end{cases}$$

The above integrals suffice for the explicit calculation of the coefficients $A_{m,n}$ for $m, n \geq 1$, as seen by Eq. (47). The integrals $J(0; W_n Z_m)$ and $J(1; W_n Z_m)$ are given by

$$J(0; W_n Z_m) = \begin{cases} \dfrac{\mu_0^2}{2 k_m^3}, & m = n \geq 1, \\ -\dfrac{k_n^{-1} \mu_0^2 + k_n - k_n \left[ Z_n Z_m \right]_{-h}}{k_n^2 - k_m^2}, & m, n \geq 1, \ m \neq n, \end{cases}$$

$$J(1; W_n Z_m) = \begin{cases} \dfrac{-\mu_0 + H(\mu_0^2 - k_m^2)}{4 k_m^3}, & m = n \geq 1, \\ -\dfrac{k_n^{-1} \mu_0 (H \mu_0 - 1) + k_n H}{k_n^2 - k_m^2}, & m, n \geq 1, \ m \neq n. \end{cases}$$

Using the three integrals given above, and the boundary values $[Z_m]_{-h}$, given by Eq. (B.1) of Appendix B, we can calculate the coefficients $\boldsymbol{B}_{m,n}$ for $m, n \geq 1$; see Eq. (48b). The needed formulae for the calculation of all coefficients are provided in Appendix B.

## 4.3 Semi-explicit calculation of local wavenumbers $k_n(\boldsymbol{x}, t)$

As explained in Subsections 4.1 and 4.2, all coefficients $A_{m,n}$, $\boldsymbol{B}_{m,n}$, $C_{m,n}$, Eqs. (18), are expressed in closed-form in terms of $H(\boldsymbol{x}, t) = \eta(\boldsymbol{x}, t) + h(\boldsymbol{x})$ and $k_n(\boldsymbol{x}, t)$ (and the fixed, auxiliary constants $\mu_0$ and $h_0$). Since the substrate problem (17) is solved by assuming that $\eta(\boldsymbol{x}, t)$ is known (either from initial data or from the previous time step), the only element which requires numerical treatment is the calculation of the roots $k_n(\boldsymbol{x}, t)$ of the transcendental equations (10). This is apparently a trivial numerical solution, which can be performed by using Newton-Raphson method. However, the requirement of very high accuracy (in order to transfer this accuracy to the coefficients), in conjunction with the fact that such a Newton-Raphson scheme would be applied about $10^{12}$ times, for demanding long-time simulations, call for fast, robust and very accurate calculations of $k_n(\boldsymbol{x},t)$. In response to these requirements, we have developed a semi-explicit solution method, providing machine-accuracy evaluations of $k_n(\boldsymbol{x},t)$ with no more than 3 iterations. The corresponding solver is given in Appendix C. A detailed description of its derivation and validation can be found in [62].

### 4.4 Significance of the analytic calculation of the coefficients

Invoking analytic expressions for the computation of the coefficients $A_{m,n}$, $\boldsymbol{B}_{m,n}$, $C_{m,n}$ allows us to fully exploit the dimensional reduction of the physical problem, enabled by the HCMT. No discretization or other numerical operations are ever employed in the $z$-direction, a feature lost if we perform numerical integration for the computation of the vertical integrals. Hence, this trait allows us not only to minimize numerical errors regarding the computation of $A_{m,n}$, $\boldsymbol{B}_{m,n}$, $C_{m,n}$, but also to greatly enhance the time-efficiency of the method. To illustrate further this feature, we compare the CPU-time needed for the computation of the coefficients by using analytic formulae against numerical integration, for a certain configuration of a physical problem. For this purpose, we consider the instantaneous geometric configuration corresponding to the problem of propagation of a solitary wave over a flat bottom with an undulating patch, Fig. 1. This problem is discussed in detail later on, in Subsection 6.4.

The horizontal extent of the computational domain is $1200\,m$ and the depth, in the flat-bottom part of the domain, is $h_0 = 1\,m$. The sinusoidal patch of the bottom is defined by Eq. (62), in Subsection 6.4, for the case of $N = 8$ undulations. The free-surface profile is a realistic one, corresponding to a specific time instant, as obtained in the course of our computations. The horizontal discretization involves in total $N_X = 12001$ points, while the number of modes used is $N_{tot} = 8$, that is, use is made of the modes -2, -1, 0, and five evanescent modes. Accordingly, each matrix $(A_{m,n})$, $(\boldsymbol{B}_{m,n})$, $(C_{m,n})$ contains 64 elements (some of which are zero).

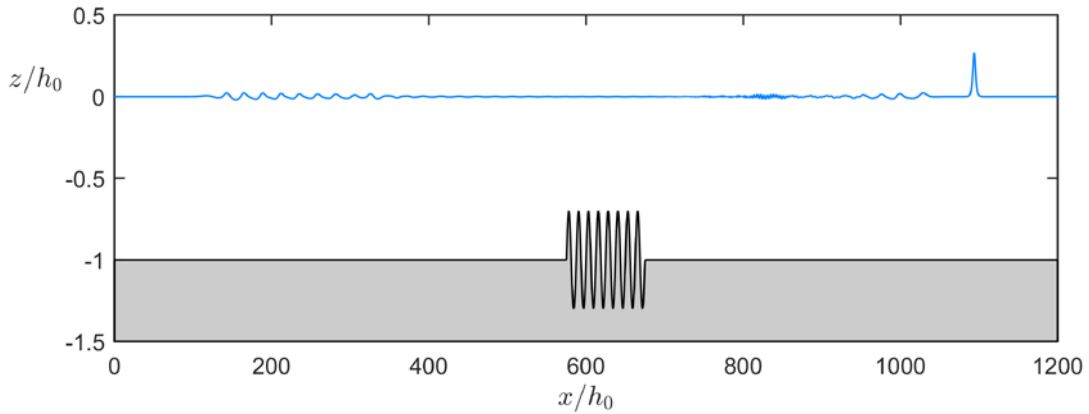

**Fig. 1.** Geometric configuration for the comparison of CPU-time between analytic calculation and numerical integration for the coefficients $A_{m,n}$, $\boldsymbol{B}_{m,n}$, $C_{m,n}$

For an accurate numerical integration of the vertical integrals, we first estimate the number of integration points needed. Taking into account the structure of the vertical functions (cf. Eq. (9c)), we see that the function $Z_n$ ($n \geq 1$) exhibits $n/2$ cycles ($n$ zeros) in the open interval $(-h, \eta)$. Thus, in the studied case, the most oscillating vertical function is $Z_5$. Hence, in order to have $8, 16, 24, 32$ number of integration points per cycle of $Z_5$, we employ, respectively, $N_Z = 21, 41, 61, 81$ vertical integration points. Simpson's rule is utilized for the numerical integration. In Table 1, we compare the CPU-time needed to compute numerically the coefficients $\{A_{m,n}\}$, $\{B_{m,n}\}$, $\{C_{m,n}\}$ [3] with the CPU-time needed to calculate *all the matrix coefficients analytically*. The relative time overhead, due to numerical integration, is calculated by the formula

$$T_{\text{rel}}[W] = (T_{\text{num}}[W] - T_{\text{anal}})/T_{\text{anal}}, \qquad \text{with } W = \{A_{m,n}\}, \{B_{m,n}\}, \{C_{m,n}\},$$

where $T_{\text{anal}}$ denotes the CPU-time needed to calculate all matrix coefficients together using the analytic expressions, and $T_{\text{num}}[W]$ denotes the CPU-time needed to calculate only the matrix coefficients $W$ by means of numerical integration. In addition, in Table 1 we also present the maximum absolute error for each group $\{A_{m,n}\}$, $\{B_{m,n}\}$, $\{C_{m,n}\}$ of coefficients over the whole computational domain, as well as their respective $L^2$-error. This table highlights two significant deficiencies of numerical integration for the coefficients:

(i) Numerical integration takes 43 (for $N_Z = 21$) up to 162 (for $N_Z = 81$) times more CPU-time than the analytic calculations. This strong CPU-time overhead deteriorates the performance of the method, rendering it unsuitable to treat extended domains for appreciable time, if one uses numerical integration for the calculation of the coefficients.

(ii) The accuracy of the evaluation of the coefficients by numerical integration is not satisfactory, especially for the coefficients $\{C_{m,n}\}$, even when a high number of integration points are used.

To further illustrate the significance of the analytic calculation concerning time efficiency of the method, we mention that, for the specific configuration shown in Figure 1, the CPU-time needed for the numerical calculation of the coefficients ranges from 5 (for $N_Z = 21$) to 21.5 (for $N_Z = 81$) times more than the time needed for solving the linear system, Eqs. (17a,b), using LU-factorization. Note that the solution of this linear system is the most time-consuming consuming component (per time step) of the whole solution procedure, when analytical computations are adopted [4].

---

[3] The notation $\{A_{m,n}\}$ means the totality of (non-zero) $A$-coefficients used in the numerical calculations, and similarly for $\{B_{m,n}\}$ and $\{C_{m,n}\}$.

[4] The computations were performed by using one core of a processor IntelCore i5-421U, 1.70GHz.

**Table 1:** Comparison of CPU-time needed for numerical integration of the matrix coefficients versus analytic calculations, together with error estimates for the numerical integrations.

| $N_Z$ | 21 | 41 | 61 | 81 |
|---|---|---|---|---|
| $T_{\text{rel}}[A_{m,n}]$ | 7.6 | 14.8 | 21.1 | 28.0 |
| max-error $[A_{m,n}]$ | $4.2 \cdot 10^{-5}$ | $2.4 \cdot 10^{-6}$ | $4.7 \cdot 10^{-7}$ | $1.5 \cdot 10^{-7}$ |
| $L^2$ − error $[A_{m,n}]$ | $1.7 \cdot 10^{-5}$ | $9.7 \cdot 10^{-7}$ | $1.9 \cdot 10^{-7}$ | $6.0 \cdot 10^{-8}$ |
| $T_{\text{rel}}[\boldsymbol{B}_{m,n}]$ | 12.1 | 24.4 | 34.2 | 45.9 |
| max-error $[\boldsymbol{B}_{m,n}]$ | $3.0 \cdot 10^{-4}$ | $1.5 \cdot 10^{-5}$ | $8.6 \cdot 10^{-6}$ | $2.7 \cdot 10^{-6}$ |
| $L^2$ − error $[\boldsymbol{B}_{m,n}]$ | $5.8 \cdot 10^{-5}$ | $2.9 \cdot 10^{-6}$ | $5.6 \cdot 10^{-7}$ | $1.7 \cdot 10^{-7}$ |
| $T_{\text{rel}}[C_{m,n}]$ | 22.5 | 45.5 | 67.0 | 87.0 |
| max-error $[C_{m,n}]$ | $7.2 \cdot 10^{-3}$ | $4.1 \cdot 10^{-4}$ | $8.0 \cdot 10^{-5}$ | $2.5 \cdot 10^{-5}$ |
| $L^2$ − error $[C_{m,n}]$ | $2.4 \cdot 10^{-3}$ | $1.4 \cdot 10^{-4}$ | $2.7 \cdot 10^{-5}$ | $8.5 \cdot 10^{-6}$ |
| $T_{\text{rel}}[A_{m,n}, \boldsymbol{B}_{m,n}, C_{m,n}]$ | 42.2 | 84.7 | 122.3 | 160.9 |

## 5. Numerical method and validations

### 5.1 Numerical solution of the substrate kinematical problem

In order to numerically implement HCMS, Eqs. (16)-(17), a method of computation for the free-surface modal amplitude $\varphi_{-2} = \mathcal{F}[\eta, h]\psi$ has to be established first. This is achieved by truncating and solving the coupled-mode system, Eqs. (17), for the instantaneous fluid domain, assuming that $\eta(\boldsymbol{x}, t)$ and $\psi(\boldsymbol{x}, t)$ are known (either from initial conditions or from the previous time step). If $M$ denotes the order of the last mode kept in the truncated series (6), the number of the unknown amplitudes $\{\varphi_n\}_{n=-2}^{M}$ involved in the computations are $N_{\text{tot}} = M + 3$. Equations for computing these unknown amplitudes are obtained by retaining the first $N_{\text{tot}} - 1$ differential equations of the system (17a), which, together with the truncated algebraic condition (17b), and the truncated version of the lateral boundary conditions, lead to a square linear system. The $(\boldsymbol{x}, t)$-dependent matrix coefficients $A_{m,n}$, $\boldsymbol{B}_{m,n}$ and $C_{m,n}$, $m, n = -2, \ldots, M$, are accurately and fast evaluated by means of the analytic formulae derived in section 4. The method has been implemented both for 2D and 3D fluid domains. In this paper, details concerning the numerical setup for the 2D case (one horizontal dimension) are presented. The 3D numerical implementation will be published elsewhere.

For the case of one horizontal dimension ($\partial_{x_1} = \partial_x, \partial_{x_2} = 0$), the truncated version of system (17) takes the form

$$\sum_{n=-2}^{M} \left( A_{m,n} \partial_x^2 + B_{mn} \partial_x + C_{m,n} \right) \varphi_n = 0, \quad x \in X, \quad m = -2,...,M-1, \tag{52a}$$

$$\sum_{n=-2}^{M} \varphi_n = \psi, \quad x \in X. \tag{52b}$$

This system is solved by the finite-difference (FD) method, using a grid of $N_X$ horizontal points, $\{x_i\}_{i=1}^{N_X}$, uniformly distributed on $X = [a,b]$. The first and second horizontal derivatives of $\varphi_n$ are approximated, at each $x_i$, by fourth-order central differences, defined by the equations:

$$(\partial_x u)^i = \frac{1}{12\Delta x}(u^{i-2} - 8u^{i-1} + 8u^{i+1} - u^{i+2}), \tag{53a}$$

$$(\partial_{x_i}^2 u)^i = \frac{1}{12\Delta x^2}(-u^{i-2} + 16u^{i-1} - 30u^i + 16u^{i+1} - u^{i+2}), \tag{53b}$$

where $\Delta x = (b-a)/(N_X - 1)$ and $u^i = u(x_i)$. Eqs. (53) are applied to $\varphi_n(x_i,t)$ for $i = 3,...,N_X - 2$. At the end points of the computational grid, $i = 1, N_X$, we use one-sided fourth-order FD, while at the adjacent points, $i = 2, N_X - 1$, we use asymmetric fourth-order FD, in order to consistently maintain the fourth-order accuracy of the numerical scheme. Substituting the FD approximations (53) in Eqs. (52), we obtain a sparse $(N_{tot} N_X) \times (N_{tot} N_X)$ linear system for the values of modal amplitudes $\{\varphi_n^i\}_{n=-2}^M = \{\varphi_n(x_i,t)\}_{n=-2}^M$, $i = 1,...,N_X$. This system, whose exact form is given in Appendix D, is solved numerically by using LU decomposition.

### 5.2 Numerical solution scheme of the evolution equations

The solution of the above-described system furnishes an approximation of the free-surface modal amplitude $\varphi_{-2}^i = (\mathcal{F}[\eta, h]\psi)^i$, $i = 1,...,N_X$, which provides a closure to the Hamiltonian equations (16). The first-order spatial derivatives, appearing in Eqs. (16), are also approximated using Eqs. (53a) with the appropriate modifications at $i = 1, 2, N_X - 1, N_X$. The resulting, semi-discretized, evolution system reads as

$$(\partial_t \eta)^i = -(\partial_x \eta)^i (\partial_x \psi)^i + \left(((\partial_x \eta)^i)^2 + 1\right) \left( h_0^{-1}(\mathcal{F}[\eta, h]\psi)^i + \mu_0 \psi^i \right), \tag{54a}$$

$$(\partial_t \psi)^i = -g\eta^i - \frac{1}{2}((\partial_x \psi)^i)^2 + \frac{1}{2}\left(((\partial_x \eta)^i)^2 + 1\right)\left( h_0^{-1}(\mathcal{F}[\eta, h]\psi)^i + \mu_0 \psi^i \right)^2, \tag{54b}$$

where $(\eta^i, \psi^i) = (\eta(x_i,t), \psi(x_i,t))$, $i = 1,...,N_X$. This system is marched in time by means of the classical, fourth-order Runge-Kutta (RK) method. The total number of time steps, $N_T$, needed to perform the computations corresponding to simulation time $T_{sim}$, is given by $N_T = T_{sim} / \Delta t$, where $\Delta t$ is the time step used. In all simulations presented in Sec. 6, the time step $\Delta t$ is selected so that the Courant number to be about $0.5$. Note that the coefficients $A_{m,n}(\eta,h)$, $B_{m,n}(\eta,h)$, $C_{m,n}(\eta,h)$ and the fields $\varphi_n, \eta, \psi$ are calculated four times within each time-step of the RK solver.

For each specific simulation, initial conditions $(\eta_0, \psi_0)$ (from which the initial $\varphi_{-2}$ is calculated) are required. In general, for nonlinear water-wave problems it is difficult to obtain physically interesting initial conditions. In this work, being focused on solitary-wave problems, initial conditions are provided by using the highly accurate, iterative solver developed in [64] for the fully nonlinear solitary wave over flat bottom.

### 5.3 Numerical accuracy and convergence

Before proceeding to solving problems concerning the interaction of solitary waves with varying bathymetry, some tests of the numerical accuracy and the actual convergence rate of the RK solver will be presented. For this purpose, the simple (and completely solved) problem of the propagation of a solitary wave over flat bottom is considered. Since, in this case, the wave remains invariant with respect to a system moving with its velocity $c$, the numerical accuracy of the RK solver for the HCMS can be measured by comparing the computed wave profile and surface potential at the end of the simulation with the corresponding initial quantities shifted by $L_{end} = c T_{sim}$. More specifically, the following $L^2$ relative error index is considered:

$$E[f] = \frac{\left\| f(x, t_{end}) - f(x + L_{end}, t_0) \right\|_{L^2}}{\left\| f(x + L_{end}, t_0) \right\|_{L^2}}, \tag{55}$$

for $f = \eta, \psi$.

The first test concerns the propagation of a solitary wave of (non-dimensional) amplitude $\alpha / h_0 = 0.5$ and velocity $c / \sqrt{gh_0} \approx 1.216$, over a flat bottom $z = -h_0 = -1m$. The horizontal domain is taken to be $X / h_0 = [0, 80]$, and the solitary wave is initially centered at $x / h_0 = 30$. The wave propagates for a small duration, $T_{sim}\sqrt{gh_0} \approx 16.4$, selected so that it covers a distance $L_{end} = c T_{sim} = 20 h_0$. The Courant number is chosen as $C = c \Delta t / \Delta x = 0.1$, in order to diminish the influence of the time integration error and allow the convergence with respect to spatial discretization $\Delta x$ to be better revealed. The errors $E[\eta]$ and $E[\psi]$ are shown in Fig. 2 (left panel) for $N_{tot} = 6, 7, 8$. In all cases, the expected rate $O(\Delta x^4)$ is very well reproduced by the numerical scheme, up to the plateau limit. Further, we observe that the errors decrease as the number of modes $N_{tot}$ increases, indicating numerical convergence of the solver with respect to $N_{tot}$.

Next, in order to assess the sensitivity of the numerical scheme with respect to the Courant number, we consider the simulation of the same solitary wave by using $C = c\,\Delta t / \Delta x = 1$. As indicated in the Fig. 2 (right panel), the errors $E[\eta]$ and $E[\psi]$ are proportional to $\Delta t^4$, showing that the present scheme converges even for coarse choices of the temporal grid.

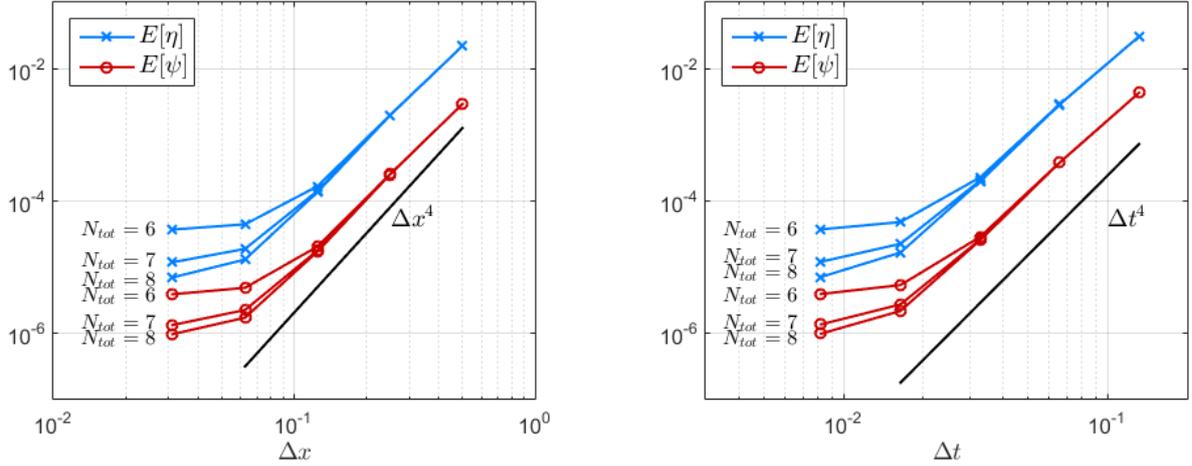

**Fig. 2.** $L^2$-error of $\eta$ and $\psi$ after a short-time propagation of a solitary wave of amplitude $\alpha / h_0 = 0.5$, for Courant number $C = 0.1$ (left panel) and Courant number $C = 1$ (right panel)

Further, in order to study the long-time performance of the RK solver for the HCMS, we simulate the propagation of the same solitary wave in a much larger horizontal domain, $X / h_0 = [0, 600]$. Now, the solitary wave is initially centered at $x / h_0 = 50$ and propagates for a duration $T\sqrt{gh_0} \approx 410.0$, selected so that the wave covers a distance of $500 h_0$. In this simulation, we use $\Delta x / h_0 = 0.125$ and a Courant number $C = 1$. Using $N_{\text{tot}} = 6$ modes, the $L^2$ errors of $\eta$ and $\psi$ are $E[\eta] = 5.7 \cdot 10^{-4}$ and $E[\psi] = 6.5 \cdot 10^{-6}$. At the end of the simulation, the solitary wave, of initial amplitude $\alpha / h_0 = 0.5$, has amplitude $\alpha_{\text{end}} / h_0 \simeq 0.499966$, and travelled exactly $500 h_0$ (hence, no phase error is present). In Fig. 3 (left panel) we compare the free-surface elevation at the end of the simulation, computed by the HCMS, with the initial free-surface elevation, obtained from [64], shifted by $500 h_0$. Clearly, the two solutions are in very good agreement. In addition, the HCMS simulation does not create a dispersive trail behind the propagating solitary wave, in contrast with many asymptotic models; see e.g. [65].

Finally, results concerning the numerical error of two fundamental conserved quantities, namely, the Hamiltonian

$$\mathcal{H}(t) = \int_X \left\{ \frac{1}{2} g \eta + \frac{1}{2} \psi \left[ -\nabla_x \eta \cdot \nabla_x \psi + \left( (\nabla_x \eta)^2 + 1 \right)\left( \varphi_{-2}\, h_0^{-1} + \mu_0 \psi \right) \right] \right\} dx \quad (56)$$

and the mass

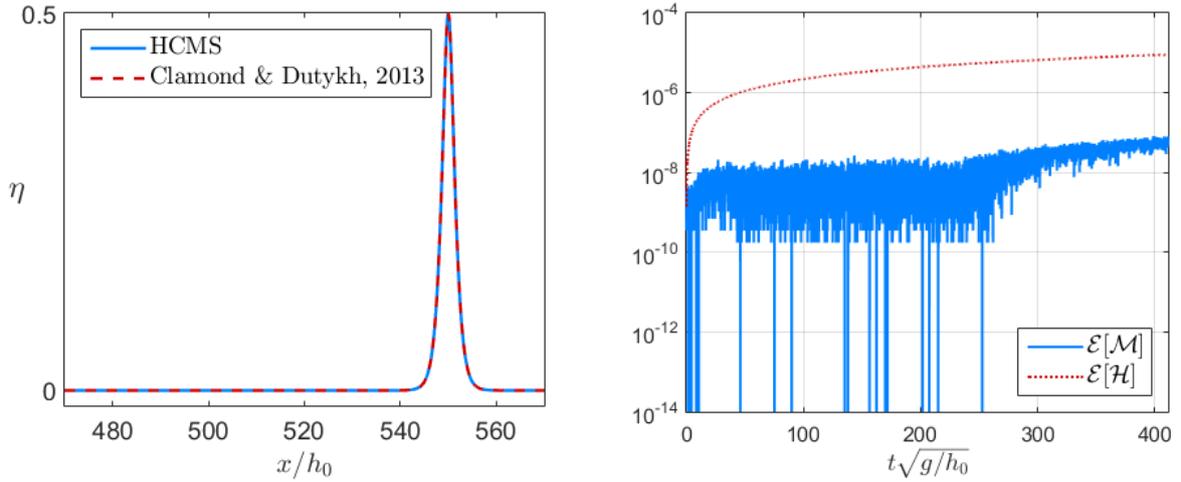

**Fig. 3.** Long-time propagation of solitary wave of initial amplitude $\alpha/h_0 = 0.5$. **Left Panel**: Comparison of free-surface elevation between HCMS simulation (solid line) and reference solution (dashed line). **Right Panel**: Conservation of mass (solid line) and the Hamiltonian (dashed line) during the simulation.

$$\mathcal{M}(t) = \int_X \eta(x,t)dx, \qquad (57)$$

are presented in the right panel of Fig. 3. The errors shown in this figure, are defined by

$$\mathcal{E}\left[\mathcal{W}(t)\right] = \left|\frac{\mathcal{W}(t) - \mathcal{W}(t_0)}{\mathcal{W}(t_0)}\right|, \qquad (58)$$

with $\mathcal{W} = \mathcal{H}, \mathcal{M}$. Both quantities are conserved up to a satisfactory accuracy during the considered long-time simulation. The relative errors $\mathcal{E}\left[\mathcal{H}(t_{end})\right]$ and $\mathcal{E}\left[\mathcal{M}(t_{end})\right]$ are $10^{-5}$ and about $10^{-7}$, respectively, showcasing the method's ability to handle long-time simulations. The slightly decreasing trend in the conservation of Hamiltonian can be attributed to the minor dissipative effects introduced by the 4th order Runge-Kutta method.

## 6. Interaction of solitary waves with bathymetry and vertical walls

In this section, numerical experiments are presented for various problems concerning the interaction of solitary waves with varying bathymetry or/and vertical walls. The first three cases refer to already studied problems, for which numerical and experimental results are available, while the remaining two cases concern problems which are studied here for the first time, as far as the authors know.

### 6.1 Reflection of solitary waves on a vertical wall

In this subsection, we study the reflection of a solitary wave propagating over a flat bottom, by a vertical wall. Such a configuration tests the ability of the HCMS to effectively simulate a fully reflecting wall condition (see Eq. (A.17)) in a strongly nonlinear case. There are several papers dealing with this problem. Fully nonlinear potential-flow simulations using BEM have been performed by several authors, e.g. [66], [67], [68] and [9]. Various approximate models have also been applied to this case, e.g. [2], [69], [70], [71]. A detailed experimental analysis of this problem has been recently published by [72], showing results for solitary waves of initial amplitude up to $\alpha/h_0 = 0.556$.

In the present simulations, the horizontal domain is taken to be $X = [a,b] = [0, 100 h_0]$, the solitary wave is initially centered in the middle of the numerical wave tank, and it propagates towards the right lateral boundary. Results for initial amplitudes up to $\alpha/h_0 = 0.66$ are obtained by using $N_{tot} = 8$ modes. The spatial discretization is $\Delta x/h_0 = 0.10$ for amplitudes $\alpha/h_0 < 0.35$, $\Delta x/h_0 = 0.05$ for amplitudes $\alpha/h_0 \in [0.35, 0.55]$, and $\Delta x/h_0 = 0.025$ for amplitudes $\alpha/h_0 > 0.55$. The time-step is selected so that the Courant constant, calculated by means of the nonlinear propagation velocity, is $C = 0.5$, for each case. Results for the normalized maximum runup on the vertical wall, defined by $R_{max}/h_0 = \max\{\eta(b,t)\}/h_0$, are shown in Fig. 4 (left panel). In the same figure (right panel) results are also presented for the normalized wall residence time $t_r/\tau$, where $\tau = \sqrt{h_0/g}$. ($t_r$ is defined as the total time for which the maximum value of $\eta$ lies on the vertical wall.) Results obtained using HCMS are compared with experimental data [72], numerical results obtained by using BEM [67], and the asymptotic formulae

$$\frac{R_{max}}{h_0} = 2\frac{\alpha}{h_0} + 0.5\left(\frac{\alpha}{h_0}\right)^2 + 0.75\left(\frac{\alpha}{h_0}\right)^3, \tag{59}$$

derived by [73], and

$$t_r/\tau = \frac{2}{\sqrt{3}}\left[\ln\left(\frac{\sqrt{3}+1}{\sqrt{3}-1}\right)(ah_0^{-1})^{-0.5} + \frac{1}{8}\ln\left(\frac{\sqrt{3}+1}{\sqrt{3}-1}\right)(ah_0^{-1})^{0.5}\right], \tag{60}$$

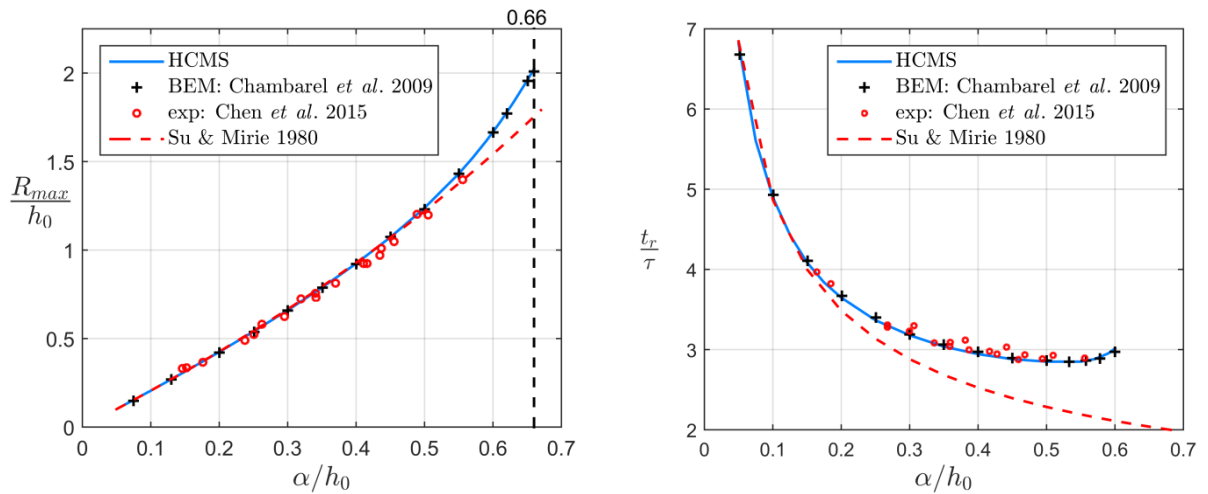

**Fig. 4.** Comparison of (left panel) normalized maximum runup and (right panel) normalized wall-residence time with other methods

derived by [66] using the perturbation theory of [73].

For all values of the initial amplitude $\alpha/h_0$, our results are identical with BEM results of [67], and in very good agreement with the experimental ones. The asymptotic Eq. (59) underestimates the maximum runup for values $\alpha/h_0 > 0.5$, while the asymptotic Eq. (60) deviates significantly from the numerical and experimental results for the wall residence time when $\alpha/h_0 > 0.2$.

For amplitudes $a/h_0 \geq 0.60$, [67] observed the formation of a residual jet during the runup (which is also the reason why their results for $t_r$ stop at $a/h_0 = 0.60$). Theoretical justification of this instability was provided in [72], Sec. 4.4. In addition, [67] showcase that for $a/h_0 \approx 0.70$ this jet is no longer a single-valued function of the horizontal position. This explains the inability of our method to simulate the last phase of the phenomenon for $a/h_0 \approx 0.70$.

In order to further compare our results with the ones provided by [67] we consider the instantaneous force acted on the vertical wall, given by

$$F_w(t) = \int_{-h_0}^{\eta(b,t)} p(b,z,t)\,dz = \int_{-h_0}^{\eta(b,t)} \left[ \partial_t \Phi + \frac{1}{2}(\nabla\Phi)^2 + gz \right]_{x=b} dz, \qquad (61)$$

where the velocity potential $\Phi(x,z,t)$ and their derivatives are computed by the series expansion, Eq. (6). Note that, this computation is very fast in the context of HCMT, since only straightforward differentiations of the terms $\varphi_n Z_n$ of the series are required([5]). Results on $F_w(t)$ in the vicinity of the maximum runup time $t_0$ are shown in Fig. 5. For the case $a/h_0 = 0.60$ the simulation is stopped when the relative error on the conservation of the Hamiltonian exceeds the value $5\times 10^{-3}$. For the case $a/h_0 = 0.7$, the present model fails to reach the maximum runup, for reasons explained above. In this case our results for the wall force stop before the maximum runup time, capturing however the maximum of the wall force (now occurring before the maximum runup) as good as the BEM results reported by [67]. Overall, our results, even for the case $a/h_0 = 0.7$, are in excellent agreement with the BEM computations, accurately capturing the variation of $F_w(t)$ during the strongly nonlinear interaction of the solitary wave with the vertical wall.

---

([5]) Note, further, that the derivatives of $\varphi_n$ have been already computed, while the derivatives of $Z_n$ are expressed in closed forms.

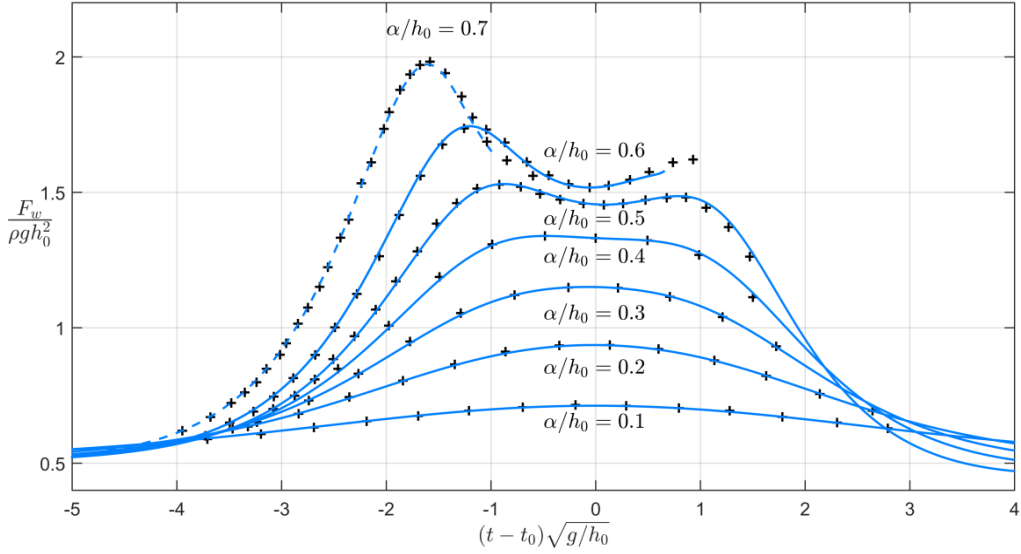

**Fig. 5.** Normalized instantaneous wall force. HCMS results (blue solid line), [67] (black crosses)

## 6.2 Shoaling of solitary waves over a plane beach

To show the applicability of HCMS to the simulation of shoaling solitary waves, we consider the benchmark test of [74], in which solitary waves of various amplitudes are generated in a laboratory wave tank and propagate towards a plane beach of a (mild) slope $1:35$, where they shoal and eventually break. See Fig. 6. The initial amplitudes of the solitary waves studied herein are $a/h_0 = 0.10, 0.15, 0.20, 0.25$. Measurements of the free-surface elevation are available at six gauges $g_0, g_1, g_3, g_5, g_7, g_9$, where the last five are located near the breaking point (see Table 2). The corresponding configuration is shown in Fig. 6.

Computations are performed by using $N_{tot} = 8$ modes, with a spatial discretization $\Delta x / h_0 = 0.02$. The time step is chosen so that the Courant constant, based on the velocity of the initial solitary wave, is $C = 0.5$. All the waves simulated here eventually break, a phenomenon HCMT cannot capture. Hence, we set a stopping criterion for the simulations to be a relative error of the Hamiltonian larger than the value $\mathcal{E}[\mathcal{H}(t)] = 10^{-3}$ (which stops the simulation a little prior to the breaking point). Comparisons of our numerical simulations with experimental data are depicted in Fig. 7. In the left panel of this figure, numerically computed free-surface elevation at the gauges $g_1, g_3, g_5, g_7, g_9$, for a solitary wave with amplitude $a/h_0 = 0.2$, are compared with experimental measurements. The computed free-surface elevations are in full agreement with the experimental data from the first four gauges. A small discrepancy appears at gauge $g_9$ (the one closest to the laboratory breaking point), with our simulation slightly overestimating the maximum wave height and speed of the wave. However, the shoaling process is overall reproduced by the HCMS with satisfactory accuracy, until the occurrence of the breaking. In the right panel of Fig. 7, results are given for the maximum amplitude of the shoaling solitary waves along the plane beach, non-dimensionalized by the local depth. For the cases $a/h_0 = 0.15,\ 0.20, 0.25$ our method is in perfect agreement with the experimental data presented in [74]. For $a/h_0 = 0.10$ our method underestimates the relative amplitude of the wave compared to the experimental results. However similar discrepancies have been observed in the BEM results as well [74]. Let it be noted that for all cases, our results seem to be in better agreement than recent simulations based on approximate models; see e.g. [9], [75] and [71].

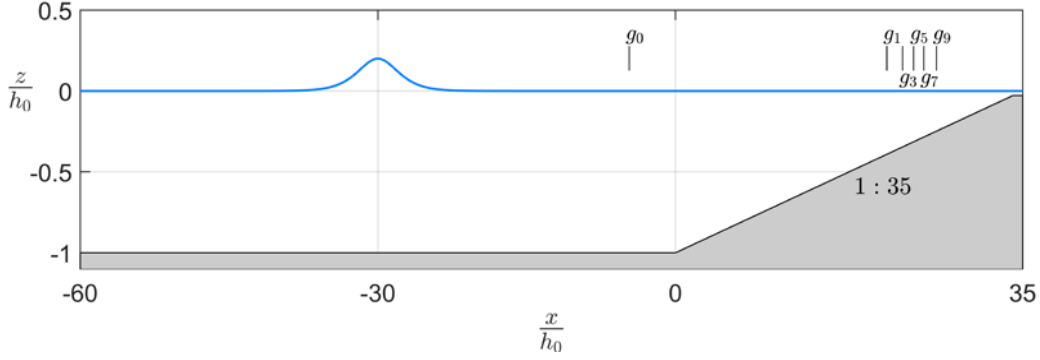

**Fig. 6.** Geometric configuration for simulations of shoaling solitary waves, with the gauges positioned for the case of $a/h_0 = 0.2$.

**Table 2:** Horizontal position of measuring gauges

| $a/h_0$ | $g_0$ | $g_1$ | $g_3$ | $g_5$ | $g_7$ | $g_9$ |
|---|---|---|---|---|---|---|
| 0.10 | -5.00 | 24.57 | 26.16 | 27.30 | 28.30 | 29.52 |
| 0.15 | -5.00 | 23.55 | 25.14 | 26.27 | 27.27 | 28.50 |
| 0.20 | -5.00 | 20.96 | 22.55 | 23.68 | 24.68 | 25.91 |
| 0.25 | -5.00 | 19.30 | 20.89 | 22.02 | 23.02 | 24.25 |

Finally, the horizontal fluid velocity field is monitored, as the wave passes through stations $g_3, g_5, g_7$, for the case of initial amplitude $a/h_0 = 0.25$. Three snapshots of these velocity fields are presented in Fig.8. To the best of our knowledge, such kind of results have not been calculated by other authors; they are presented here for the first time. We observe that the horizontal velocity in the vicinity of the crest rapidly increases as the wave travels over the slope. For example, when the maximum free-surface elevation is over the gauges $g_3, g_5, g_7$ (the three cases shown in Fig. 8), the maximum horizontal velocity is $U_{x,\max}/\sqrt{g h_0} = 0.535$, $0.611$ and $0.717$, at horizontal positions $x_{\max}/h_0 = 20.94, 22.08, 23.06$, respectively, being slightly ahead of the corresponding position of the maximum elevation.

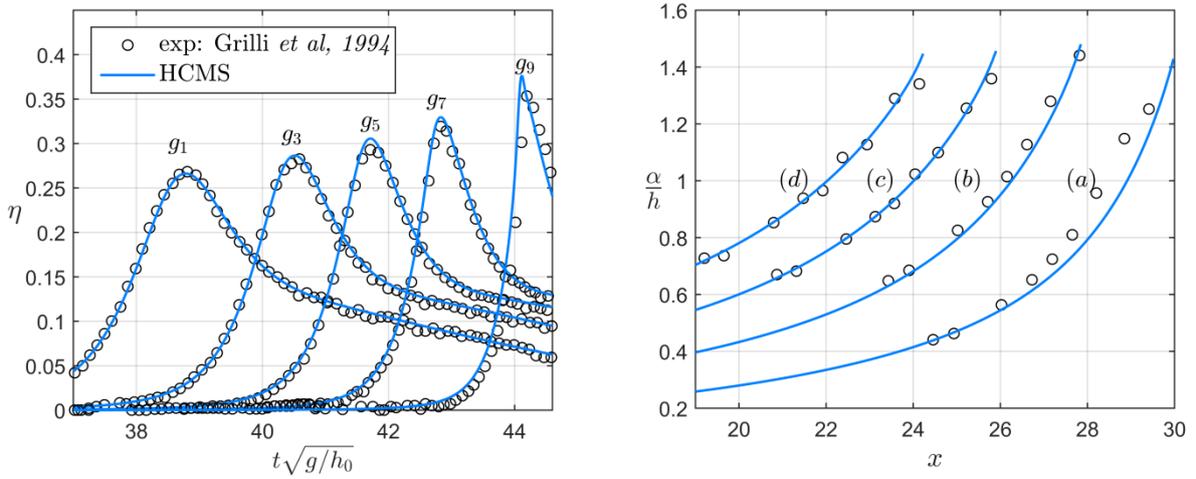

**Fig. 7.** Right figure: Comparison of free-surface elevation history at the gauges of Table 1, between experimental results from [74] (circles) with our simulation (solid line). Left figure: Comparison of relative amplitude between HCMS (solid line) and experimental data (circles) for solitary waves of starting relative amplitude (a) 0.10 (b) 0.15 (c) 0.20 (d) 0.25.

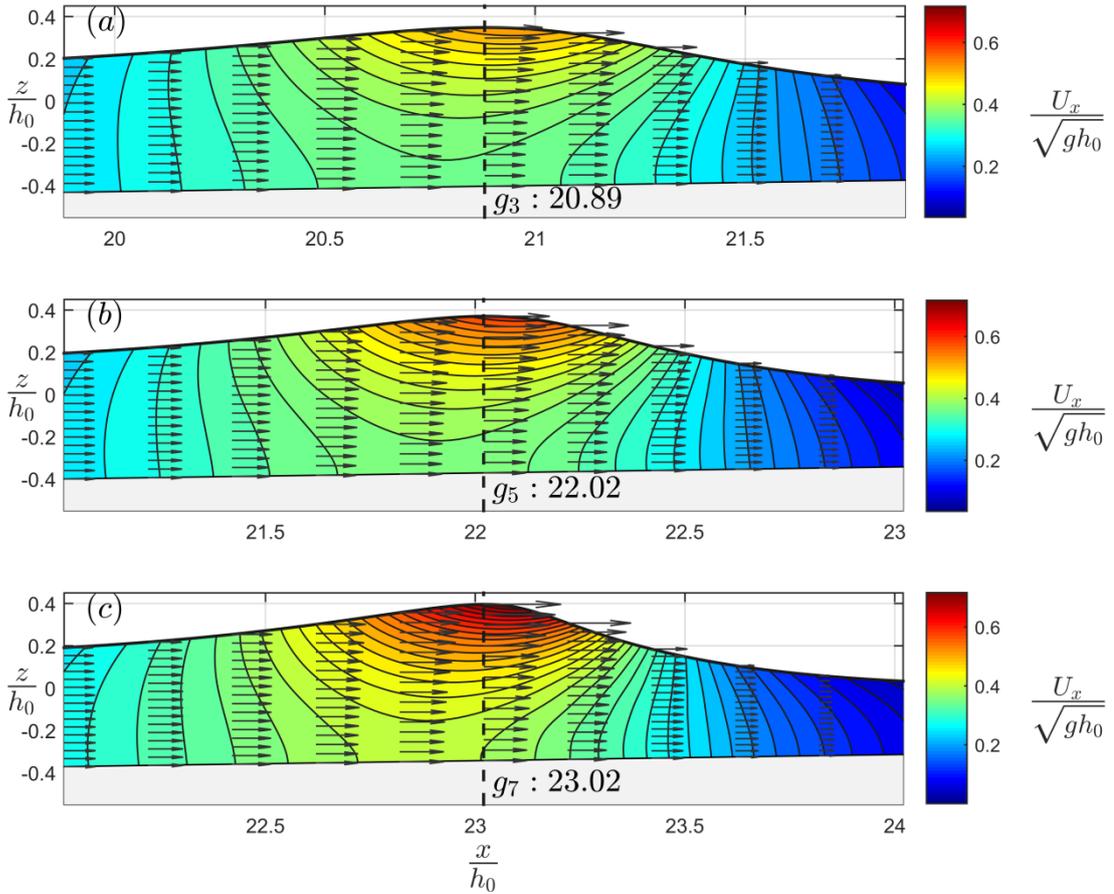

**Fig. 8.** Horizontal velocity field of a shoaling solitary wave having initial amplitude $a/h_0 = 0.25$, as it passes through stations (a): $g_3$, (b): $g_5$, (c): $g_7$.

## 6.3 Reflection of shoaling solitary waves on a vertical wall at the end of a sloping beach

In this numerical test, we simulate solitary waves that shoal, climbing up a plane beach of slope $1/50$, and reflect on a vertical wall without breaking. Experimental measurements of the local free surface elevation during this complex process have been used for the validation of several *Boussinesq type* equations [76], [77], [78]. As reported in these works, such model equations, being weakly nonlinear and weakly dispersive, capture the phenomenon with moderate accuracy, especially when the amplitude of the incident solitary wave is significant; see e.g. Section 5.2 of [77]. Recently, [71] demonstrated that the fully nonlinear *Serre-Green-Nagdhi* (SGN) equations simulate this experiment with higher accuracy than weakly nonlinear models. As will be demonstrated below, our method, being fully nonlinear, captures nicely this phenomenon, as expected.

The horizontal computational domain extends from $x = -100\,m$ to $x = 20\,m$, where vertical impermeable walls are located. The depth, for $x \in [-100, 0]$ is constant, $h_0 = 0.7$ m, and the plane beach starts at $x = 0\,m$ and ends at $x = 20\,m$. The initial solitary wave is centered at $x = -30\,m$; see Fig. 9. We have considered both available cases of this experiment, corresponding to solitary waves of amplitudes $\alpha_0 = 0.07\,m$ and $0.12\,m$, and velocities $c = 2.7472\,m/s$ and $c = 2.8337\,m/s$, respectively. For the numerical solution of HCMS, we have used $N_{tot} = 7$ modes, and a spatio-temporal discretization $\Delta x = 0.08\,m$ and $\Delta t = 0.015\,s$. Comparisons of our computations with experimental measurements are shown in Figures 10 and 11 for the cases $\alpha_0 = 0.07\,m$ and $0.12\,m$, respectively. In the first case, the agreement is perfect, that is, the fully nonlinear potential HCMT completely captures the phenomenon. Thus, the 8% overprediction of the reflected wave at station #3 by the Boussinesq equations of [77] (p. 152) cannot be attributed to the lack of viscosity, as conjectured by the authors of the above paper, but to the nonlinear effects. The agreement is also very good for the second case, $\alpha_0 = 0.12\,m$, with a slight overprediction (less than 10%) of the reflected wave (second peak). Note that the corresponding overprediction by the aforementioned Boussinesq equations is 38%. The discrepancy between simulation and experimental measurements, appearing left and right of the two peaks in this case, is also present when the simulations are performed by other models, e.g. the fully nonlinear SGN equations; see e.g. [71], Fig. 11. The above results demonstrate the ability of the present method to accurately simulate the nonlinear shoaling and reflection processes.

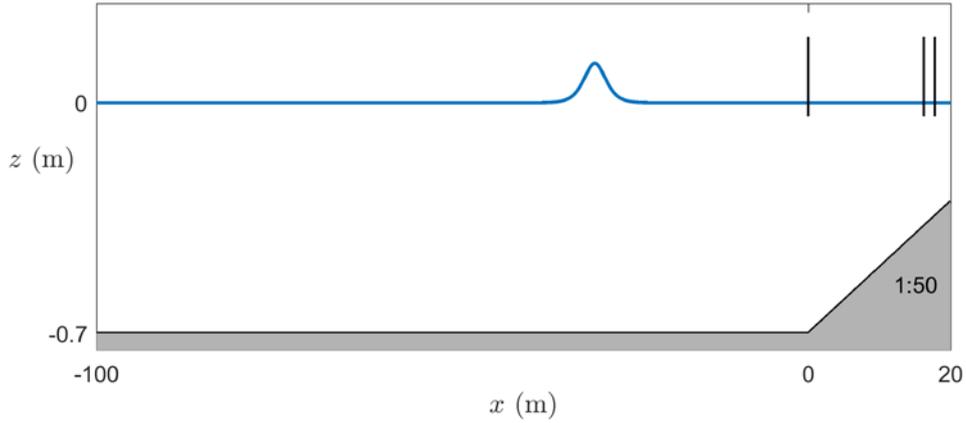

**Fig. 9.** Initial free surface and bathymetry for the reflection of shoaling solitary waves. Vertical black line segments correspond to the measuring stations #1, #2 and #3, located at $x = 0$, 16.25 and 17.75 $m$.

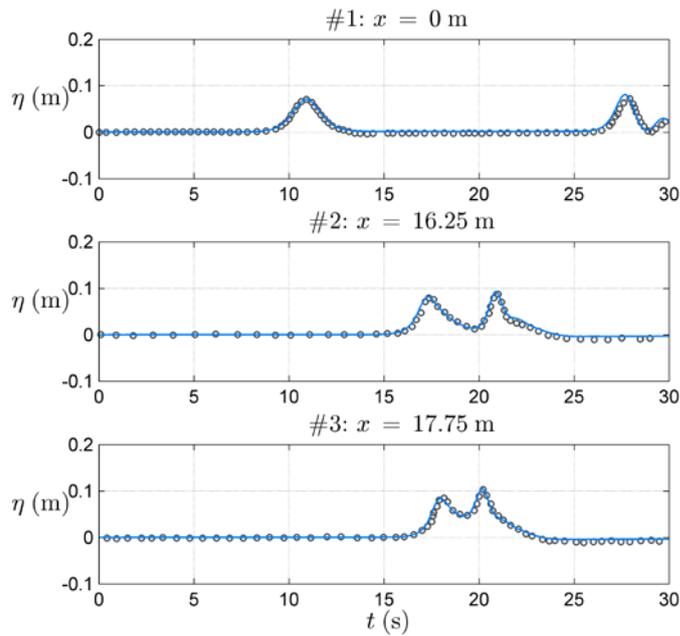

**Fig. 10.** Time series of the free-surface elevation, at gauges #1, #2, #3, as indicated of the header of each subplot, for the problem of shoaling and reflection of a solitary wave with initial amplitude $\alpha_0 = 0.07\,m$. Numerical solution is shown with a solid (light blue) line, and experimental data with circles.

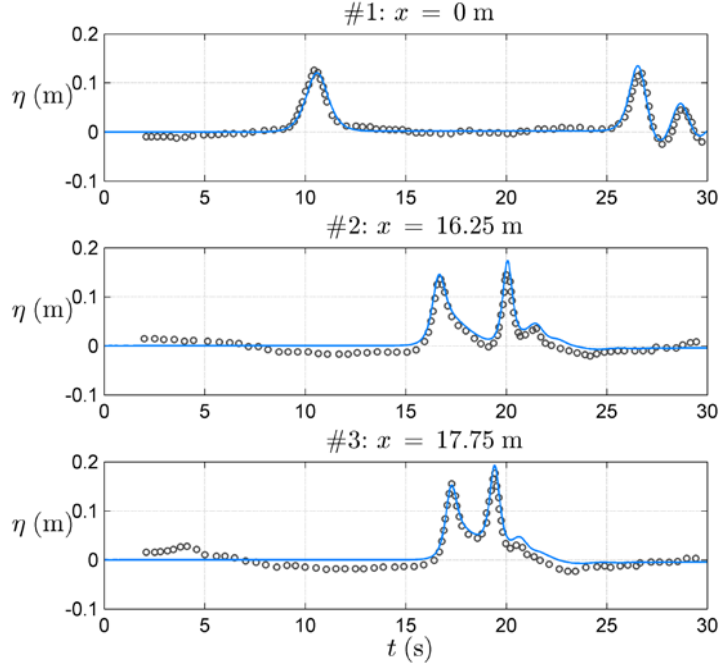

**Fig. 11.** The same as in Fig. 10, for a solitary wave with initial amplitude $\alpha_0 = 0.12\,m$.

### 6.4 Propagation of a solitary wave over a sinusoidal patch

Having established the efficiency and the accuracy of the HCMS by means of the three previous case studies, we now proceed with its application to new interesting problems, which are investigated herein apparently for the first time. In this subsection we study the transformation of a solitary wave of initial amplitude $a/h_0 = 0.30$ and velocity $c/\sqrt{gh_0} \approx 1.1375$ as it moves over a seabed with a sinusoidal patch of finite extent; see Fig. 12. The sinusoidal patch has an horizontal length of $100\,m$ and an amplitude of $0.3\,m$. The solitary wave is initially centered at $x = 500\,m$, within a computational domain $X = [0, 1200\,m]$. We study four different cases, corresponding to four different periods of the sinusoidal patch. The bathymetry is given by the formula (in meters)

$$h(x) = \begin{cases} 1 - 0.3\sin\left[2\pi(x - 575)N/100\right], & x \in [575, 675], \\ h_0 = 1, & \text{otherwise}, \end{cases} \quad (62)$$

where $N = 2, 4, 6, 8$ denotes the number of ripples in the sinusoidal patch. The configuration of the problem for $N = 4$ is presented in Fig. 12.

The horizontal grid consists of $N_X = 12001$ points, and the total number of modes used is $N_{tot} = 6$. The time-step is selected so that the Courant number is $C = 0.5$ initially. During the simulations, the mass and the Hamiltonian conservation errors are never larger than $\mathcal{E}[\mathcal{M}(t)] = 10^{-5}$ and $\mathcal{E}[\mathcal{H}(t)] = 3\cdot 10^{-5}$, respectively. The simulations discussed herein are stopped at time $t_{end}$, defined so that $ct_{end} = 670m$ ([6]). At the end of each simulation, the solitary wave will have evolved into a complex wave pattern, including waves of different spatio-temporal scales. See Fig. 13. The main constitutes of this complex wave pattern are: (i) a frontal, lower-amplitude, solitary-like wave moving forward, (ii) a complex, small-amplitude, wave-train trailing the solitary wave, and (iii) a reflected (by the sinusoidal bottom), back-propagated, wave-train. In Fig. 13, this wave pattern is shown for the cases $N = 8$ and $2$.

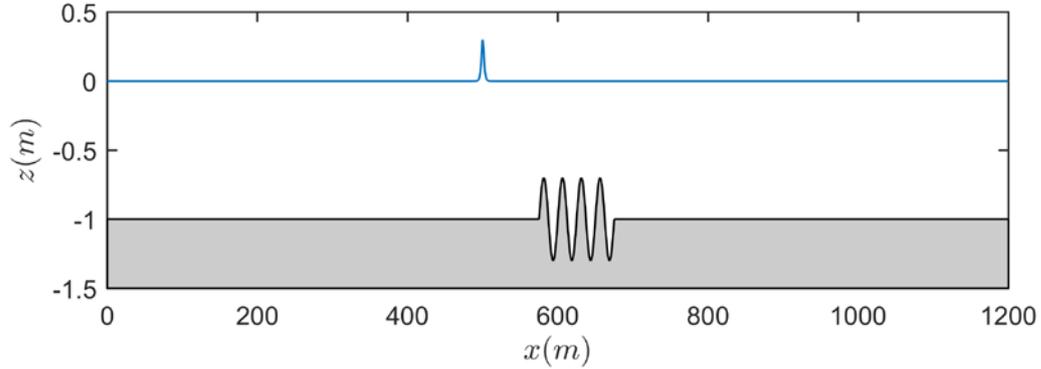

**Fig. 12.** Initial Configuration for the case of $N = 4$ ripples of the sinusoidal patch.

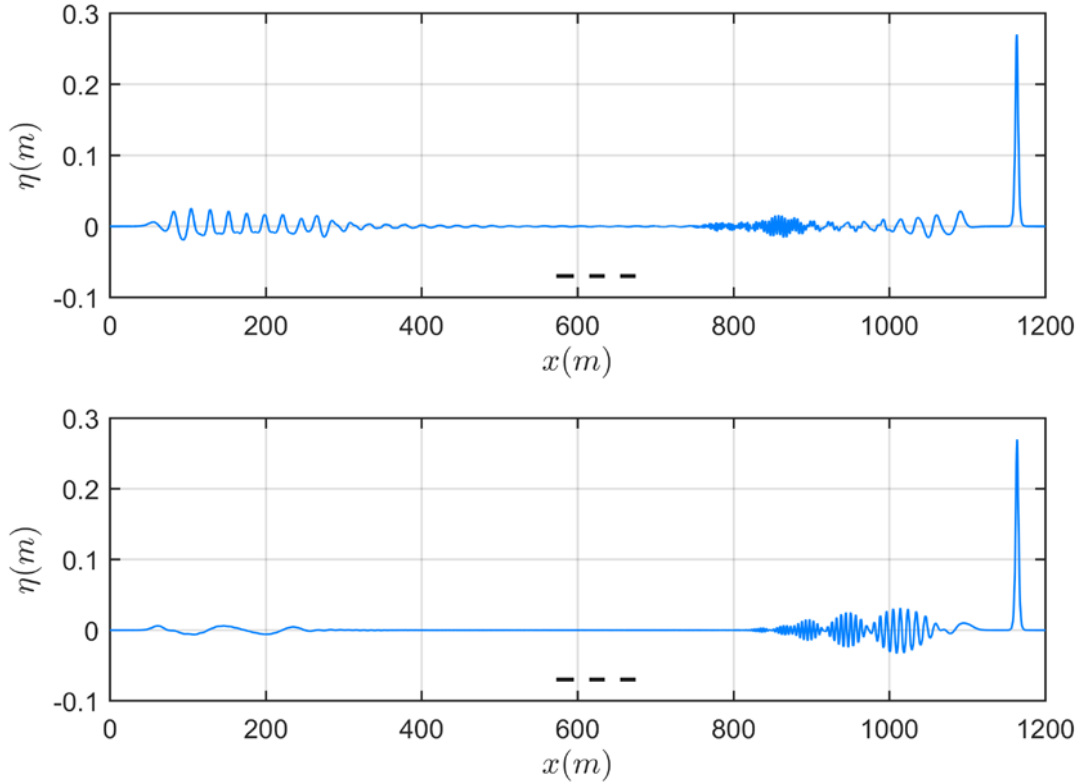

**Fig. 13.** The structure of the free-surface elevation at $t_{end}$ for $N = 8$ (upper panel), and $N = 2$ (lower panel). The position of the patch is designated by the dashed line.

---

([6]) The code can continue, simulating also the reflection of the transmitted wave system on the right wall.

In addition, during the simulation, a quasi-standing (leaking) wave system is developed in the region over the sinusoidal patch, whose energy is slowly escaping from that region, moving both in the forward and the backward direction. This phenomenon, which is more pronounced as the number $N$ of ripples increases, can be clearly observed in [video 1](video 1) (for the case $N = 8$), provided as electronic supplementary material.

We focus now on how the initial energy is separated into transmitted and reflected parts when the solitary wave has travelled past the sinusoidal patch. More precisely, we calculate the evolution of energy inside the following three distinct regions. The first one engulfs the reflected wave train and extends from the left lateral boundary $x = 0\,m$ up to $x = 574\,m$ (1 meter before the left boundary of the sinusoidal patch). The second region is the one where the transmitted wave system resides, and extends from $x = 676\,m$ up to $x = 1200\,m$. The third region, extends above the sinusoidal patch, $x \in [574\,m, 676\,m]$. The energy contained in the last region is called subsequently residual energy. The distribution of energy in the three regions at the end of each simulation is presented in Table 3. As can be seen, a relatively small amount of energy is reflected, which increases with the frequency of the sinusoidal patch. The residual energy is essentially zero for all cases (at the end of the simulation). However, as already noted, during the simulations, leaking waves are observed slowly escaping from that region, becoming negligible as time passes. Fig. 14 displays the time evolution of reflected, transmit-ted and residual energy, for the case $N = 8$, from time $t_1 = 56.14\,\text{sec}$ (shortly after the leading solitary-like wave passes the patch) up to the end of the simulation.

**Table 3:** Reflected, transmitted, and residual energy percentages at $t_{end}$

| $N =$ | 2 | 4 | 6 | 8 |
|---|---|---|---|---|
| Reflected energy | 1.14% | 3.66% | 6.41% | 9.40% |
| Transmitted energy | 98.86% | 96.34% | 93.59% | 90.59% |
| Residual energy | 0.00% | 0.00% | 0.00% | 0.01% |

As an additional validation test of the HCMT and its numerical solver for this problem, we simulate the reverse evolution for the case $N = 8$. More precisely, we use as initial data the fields $\eta_0 = \eta(x, t_{end})$ and $\psi_0 = -\psi(x, t_{end})$, and redo the simulation. The dynamical system, being a Hamiltonian one, is reversible in time, and this property is nicely reproduced in numerical simulation. The complex wave pattern, appearing at $t_{end}$, evolving backward in time, merges to form the initial solitary wave, now propagating in the opposite direction. The whole process can be viewed in [video 2](video 2), provided as electronic supplementary material. To quantify the accuracy of the reversible evolution, we compare the amplitude and the phase (position) of the initial solitary wave with the corresponding ones of the reproduced solitary wave. We report herein that the amplitude of the reproduced solitary wave differs from the initial amplitude ($a = 0.30\,m$) only by $3.6 \cdot 10^{-6}\,m$ (being lower), while it is centered exactly at $x = 500m$, which means that no phase error exists. The $L^2$-relative error of the free-surface elevation field, at $t = 0$, between the two simulations, defined similarly as in Eq. (55), is $E[\eta] = 9.8 \cdot 10^{-5}$.

**6.5 Transformation of a solitary wave over a 3D bathymetry with banks and trenches**

Our final numerical experiment refers to a complex 3D configuration, with a seabed consisting of trenches and banks, over which an initially 2D solitary wave propagates, focusing over banks and disintegrates to a complicated dispersive pattern. The reflection of the resulting 3D wave pattern from a vertical wall is also simulated. The goal of this case study is two-fold. First, to show the ability of HCMS to treat 3D fully nonlinear problems and, second, to initiate the investigation of an interesting problem, with a rich physical content, hoping that it will motivate further similar studies, either experimental or numerical.

The initial free-surface elevation for this problem is that of a solitary wave of amplitude $\alpha/h_0 = 0.25$, moving along the $x_1$–direction, and extended along the transverse $x_2$–direction of a computational domain $(x_1, x_2) \in X = [0, 200\,m] \times [0, 50\,m]$. In the vicinity of the initial position of the solitary wave, bathymetry is flat, with depth $h_0 = 1m$, being smoothly transformed to a system of parallel banks (of depth $h_0 = 1m$) and trenches (with maximum depth $h_{tr} = 3m$); see Fig. 15. The computational domain is symmetric with respect to the vertical plane containing the $x_1$ – axis. Exact mathematical description of the bathymetry is given in Appendix E.

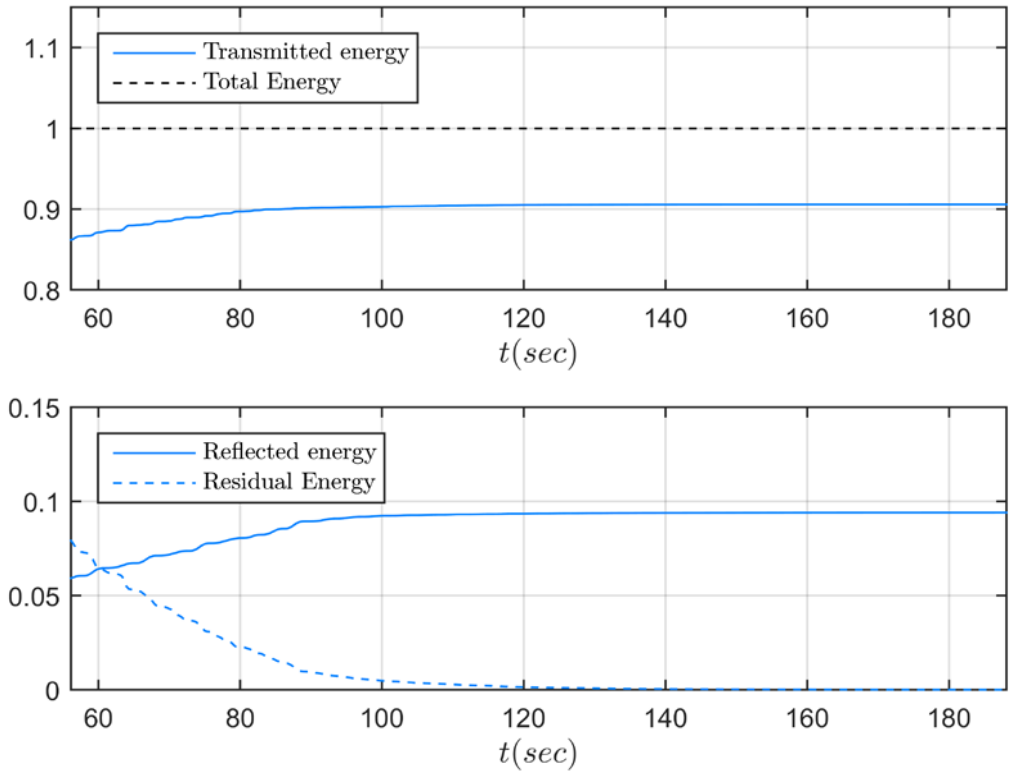

**Fig 14.** Evolution of transmitted and total energy (upper panel), and of reflected and residual energy (lower panel), for the case $N = 8$, during the period $[t_1, t_{end}]$.

The solitary wave is initially centered at $x_1 = 30\,m$ (see Fig. 16, upper panel). The simulation is performed up to time $t_{end} = 91\,\sec$, by using a spatial discretization of $N_{X_1} \times N_{X_2} = 801 \times 201$ grid points, $N_{tot} = 6$ modes, and a time-step such that the Courant constant is $C = 0.5$ at the initial time. All lateral boundaries are taken to be vertical walls.

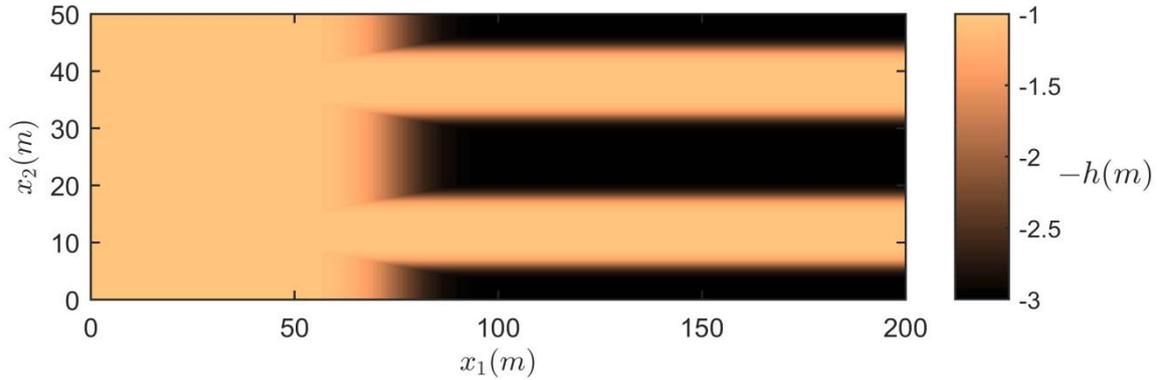

**Fig. 15.** Bottom topography of the problem

As the solitary wave propagates over the varying bathymetry, it undergoes a complex transformation phenomenon, consisting of several phases. In a first phase, a pronounced focusing is observed along the banks; see lower panel of Fig. 16. The maximum amplitude of the free-surface elevation surpasses that of the incident solitary wave, reaching the value $0.2976\,m$ (19% larger than the initial amplitude $a/h_0 = 0.25$). In a second phase, the amplitude of the leading waves decreases, due to energy transfer into the complicated dispersive trail; Fig. 17 (upper panel). Backward reflected waves, due to bottom variability, although visible, are negligible in comparison with the propagating wave train, as expected because the descent of the bottom depth is smooth and not abrupt. As the wave system evolves further, interference phenomena appear in the rear; Fig. 17 (upper panel). During the reflection on the right wall (positioned at $x_1 = 200\,m$), the maximum runup is $R/h_0 = 0.1830$. After the reflection (Fig. 17, lower panel), the free surface becomes quite complicated, with a slight focusing along the banks. At the final simulation phase, and as the front of the back-reflected wave approaches the flat region (at the left part of the computational domain), a well-formed, quasi 2D, solitary-like wave emerges; see Fig. 17 (lower panel). The full simulation, including all phases described above, can be seen in video 3, included as electronic supplementary material.

In Fig. 18, we present the horizontal velocity in the $x_1$-direction along two cuts, one in the middle of a bank (upper panel) and one in the middle of a trench (lower panel), at some intermediate time $t = 35.76\,\sec$. As one can expect, the $U_{X_1}$ horizontal fluid velocity over the bank is larger than the velocity over the trench, while the variation along the $z$-axis is much more pronounced in the second case. In Fig. 19, we present the horizontal velocity in the $x_2$-direction, $U_{X_2}$, on a $x_2$-cut at $x_1 = 141.50\,m$ and $t = 35.76\,\text{sec}$. It is clearly seen that this velocity is strongly varying, being larger (in absolute value) near the edges of the banks.

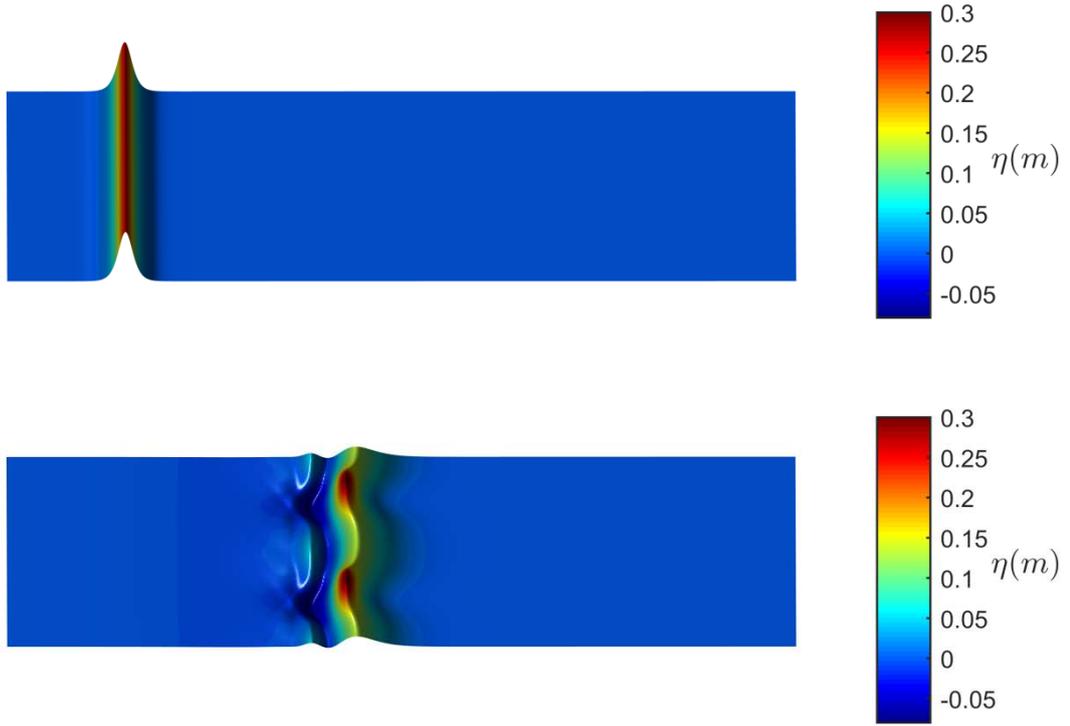

**Fig. 16.** Plot of free-surface elevation (upper panel) at $t = 0\,\text{sec}$, initiation of the simulation, and (lower panel) at $t = 15.70\,\text{sec}$, time of maximum free-surface elevation

During the simulation, the error in mass and Hamiltonian conservation do not exceed the values $\mathcal{E}[\mathcal{M}(t)] = 4\cdot 10^{-4}$ and $\mathcal{E}[\mathcal{H}(t)] = 6\cdot 10^{-5}$. Also, the symmetry of the free-surface elevation with respect to the geometric symmetry plane of the domain is maintained. These results provide strong indications concerning the correctness of the simulation.

The complete dataset of this numerical simulation is available, upon request, to any researchers for comparisons and further investigation.

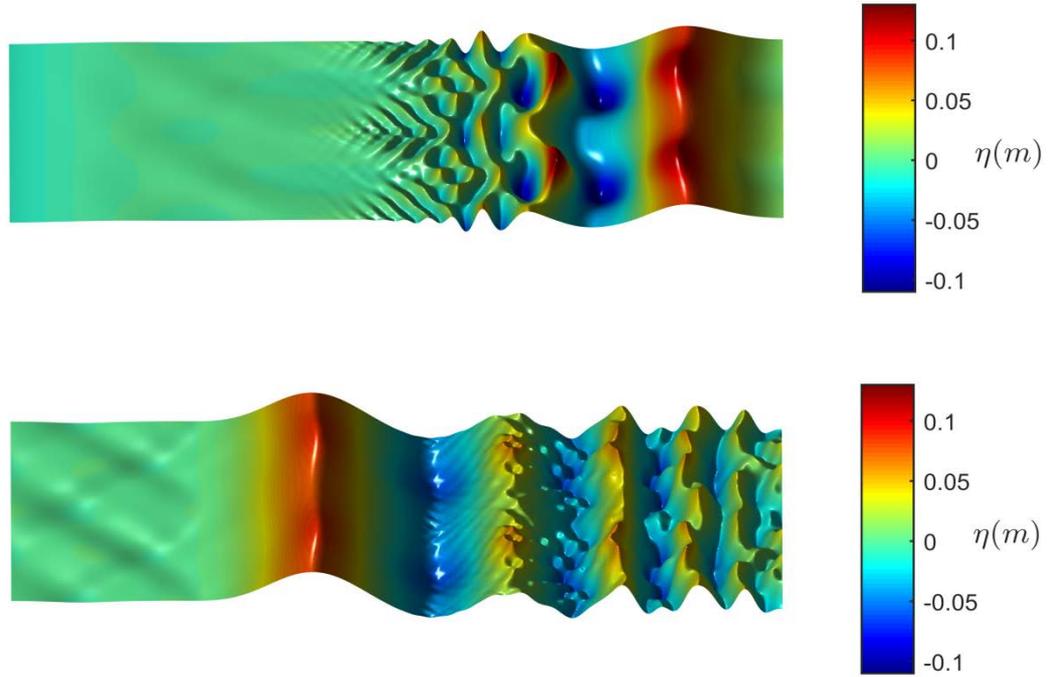

**Fig. 17.** Upper panel: Free-surface elevation at $t = 35.76\,\text{sec}$, shortly prior the reflection of the leading wave at the right wall. Lower panel: Free-surface elevation at $t_{\text{end}} = 68.73\,\text{sec}$, after the wave-train has been reflected at the right wall.

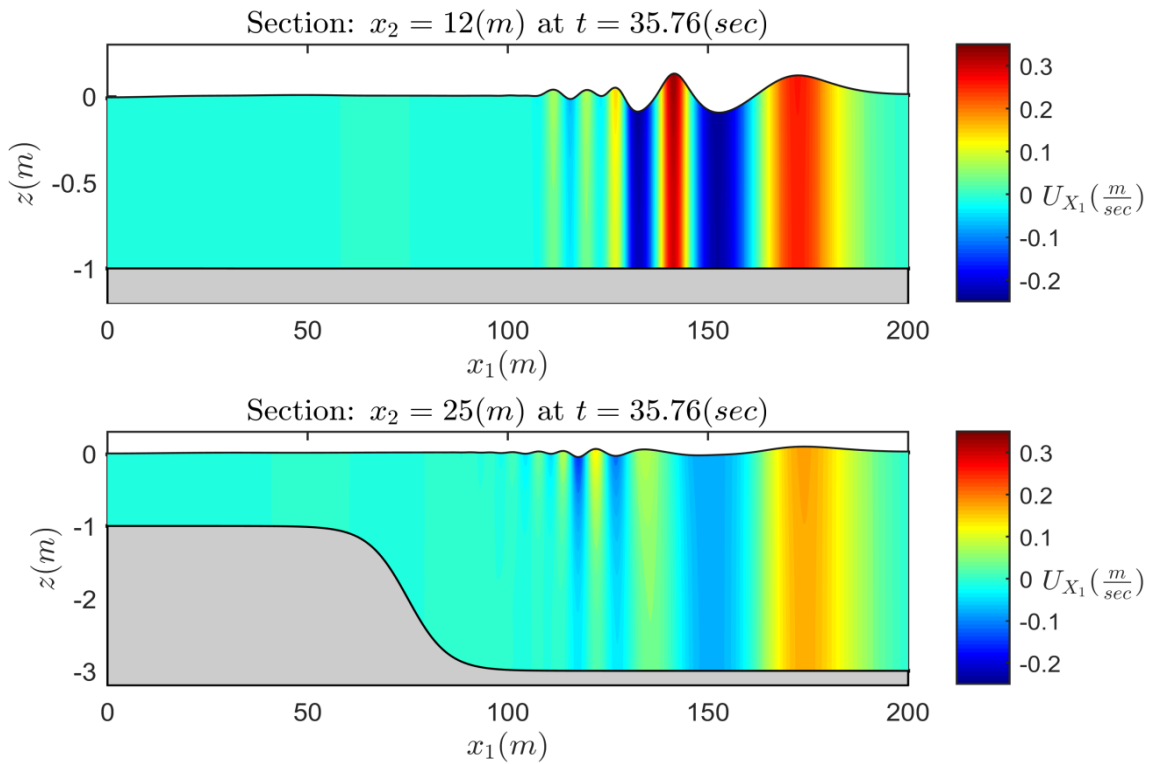

**Fig. 18.** Horizontal Velocity (in the $x_1$-direction) at time $t = 35.76\,\text{sec}$. Upper panel: over the center of one of the banks; Lower panel: over the center of the central trench.

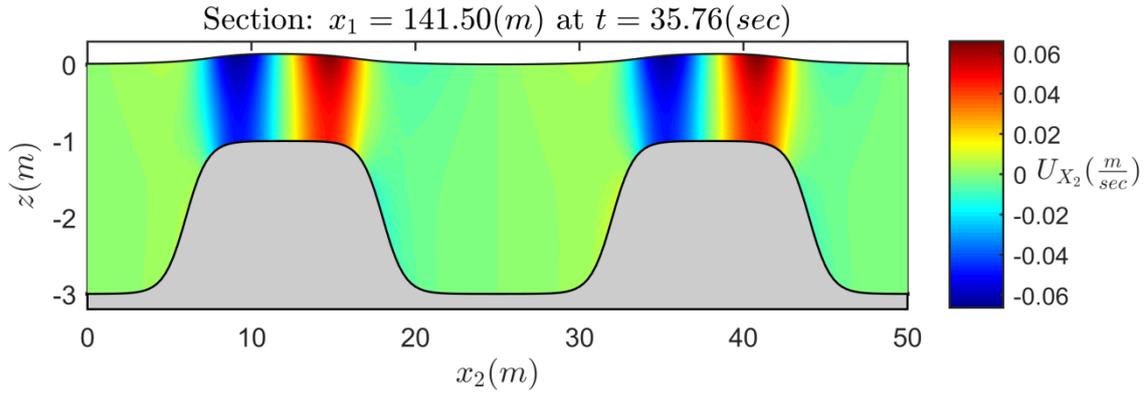

**Fig. 19.** Horizontal velocity in the $x_2$-direction at time $t = 35.76\,\text{sec}$ at $x_1 = 135.75\,m$ (position of the highest crest for that time).

## 7. Discussion and conclusions

The Hamiltonian Coupled-Mode Theory (HCMT), recently launched by [50], has been briefly described, putting emphasis, in the present paper, on its efficient numerical implementation. Applications to the study of various phenomena occurring when a solitary wave interacts with a varying bathymetry or/and a vertical wall have also been presented.

HCMT reduces the nonlinear, free-boundary problem of water waves to a system of two Hamiltonian evolution equations, coupled with an elliptic substrate problem, defined on a fixed horizontal domain. The variation of the physical domain due to the unknown free surface is embodied both in the two evolution equations and in the substrate problem, whose coefficients $A_{m,n}$, $B_{m,n}$ and $C_{m,n}$, are dependent on the instantaneous free-surface elevation. A critical part of the implementation of HCMT is the calculation of these space-time varying coefficients, which should be updated (four times) at each time step. This results in $10^8 - 10^{12}$ evaluations of these coefficients, becoming the most time-consuming element of the whole procedure, if implemented numerically. To overcome this bottleneck, analytical calculations of all coefficients have been performed, reducing their computation time to $10-15\,\%$ of the total computation time.

The present method, accelerated as described above, offers a non-perturbative, dimensionally reduced computational approach, having the same accuracy as direct numerical methods (e.g., BEM, see results in Sec.6), being significantly faster than them. In addition, the present method permits the accurate calculation of all flow fields (velocities, accelerations, pressure) throughout the whole domain, by a straightforward post processing of very low computational cost.

Further, HCMT has been applied to the study of various phenomena taking place when a solitary wave interacts with a varying bathymetry or/and a vertical wall. First, the accuracy and convergence of the numerical scheme is investigated against the steady propagation of highly accurate solitary wave solutions of Euler equations. Second, the numerical scheme was validated by comparisons against existing experimental measurements and BEM computations for three widely studied benchmark cases, involving nonlinear transformation of solitary waves (reflection and shoaling). Finally, two novel applications are presented to configurations seemingly not examined yet: a) the reflection, transmission and disintegration of a solitary wave by a finite undulating seabed patch, and b) the propagation, focusing and disintegration of a solitary wave moving over a 3D bathymetry, along troughs and banks, and the reflection of the resulting complicated wave pattern on a vertical wall. Velocity fields throughout the fluid are also presented for some of the above cases. It is hoped that the novel results presented herein will motivate experimental studies for further investigation and comparisons. Videos of the free-surface evolution for some of the studied cases are available through the electronic supplementary material.

**Acknowledgments:** The authors are grateful to Associate Professor Theodoros Gerostathis, of the Technological Educational Institute of Athens, for his assistance during the development of the code used in this work. Ch.E. Papoutsellis was supported by a PhD scholarship from NTUA.

**Appendix A: An outline of the derivation of the Hamiltonian Coupled Mode System**

The goal of this appendix is to outline the derivation of the HCMS, consisting of Eqs. (16), (17) and (18). These equations were first announced in [79], and their detailed proof was presented in [50]. Since the proof is complicated and lengthy, the present appendix would be useful even to those readers who are willing to follow the detailed proof, as well.

As stated in Sec. 2, the fields $\eta$ and $\Phi$ satisfy the NLPF problem if and only if they satisfy the variational equation (5) [55]. Further, Eq. (5), as any variational equation, is invariant under bijective, bi-continuous and bi-differentiable changes of functional variables. The functional transformation (15), defined through the exact representation (6), written here in the form

$$\Phi = \Phi[\eta, \boldsymbol{\varphi}] = \sum_{n=-2}^{\infty} \varphi_n Z_n(\eta) \equiv \boldsymbol{\varphi}^{\mathrm{T}} \boldsymbol{Z}(\eta), \quad (^7) \tag{A.1}$$

where $\boldsymbol{\varphi} = \{\varphi_m(\boldsymbol{x},t)\}_{m=-2}^{\infty}$ and $\boldsymbol{Z}(\eta) = \{Z_m(z;\eta,h)\}_{m=-2}^{\infty}$, possesses the aforementioned properties. Thus, an equivalent variational principle can be formulated with respect to $\eta$ and $\boldsymbol{\varphi} = \{\varphi_m(\boldsymbol{x},t)\}_{m=-2}^{\infty}$, by means of the transformed action functional

$$\widetilde{\mathcal{S}}[\eta, \boldsymbol{\varphi}] = \mathcal{S}[\eta, \Phi[\eta, \boldsymbol{\varphi}]] = \mathcal{S}[\eta, \boldsymbol{\varphi}^{\mathrm{T}} \boldsymbol{Z}(\eta)]. \tag{A.2}$$

The corresponding variational equation reads as

$$\delta \widetilde{\mathcal{S}}[\eta, \boldsymbol{\varphi}; \delta\eta, \delta\boldsymbol{\varphi}] = \delta_\eta \widetilde{\mathcal{S}}[\eta, \boldsymbol{\varphi}; \delta\eta] + \sum_{n=-2}^{\infty} \delta_{\varphi_m} \widetilde{\mathcal{S}}[\eta, \boldsymbol{\varphi}; \delta\varphi_m] = 0, \tag{A.3}$$

---

($^7$) $\boldsymbol{\varphi}$ and $\boldsymbol{Z}(\eta) = \boldsymbol{Z}$ are considered as column vectors, and $\boldsymbol{\varphi}^{\mathrm{T}}$ denotes the transpose of $\boldsymbol{\varphi}$.

and is equivalent with Eq. (5). Since the new functional $\widetilde{S}[\eta,\boldsymbol{\varphi}]$ is a composite one, the calculation of its functional derivatives, $\delta_\eta \widetilde{S}$ and $\delta_{\varphi_m}\widetilde{S}$, is facilitated by using the chain rule for functional derivatives; see e.g. [80], Prop. 4.1.12. As explained in detail in [50], the application of the chain rule leads to the equations

$$\delta_\eta \widetilde{S}[\eta,\boldsymbol{\varphi};\delta\eta] = \delta_\eta S[\eta,\boldsymbol{\varphi}^\mathrm{T}\mathbf{Z};\delta\eta] + \delta_\Phi S[\eta,\boldsymbol{\varphi}^\mathrm{T}\mathbf{Z};(\boldsymbol{\varphi}^\mathrm{T}\partial_\eta \mathbf{Z})\delta\eta], \tag{A.4}$$

$$\delta_{\varphi_m}\widetilde{S}[\eta,\boldsymbol{\varphi};\delta\varphi_m] = \delta_\Phi S[\eta,\boldsymbol{\varphi}^\mathrm{T}\mathbf{Z};Z_m \delta\varphi_m], \tag{A.5}$$

where $\partial_\eta \mathbf{Z}(\eta) = \{\partial_\eta Z_m(z;\eta,h)\}_{m=-2}^\infty$, and $\delta_\eta S$, $\delta_\Phi S$ are the functional derivatives of the initial functional $S[\eta,\Phi]$ with respect to $\eta,\Phi$, respectively. A convenient (for our purposes) form of the latter is given by the formulae:

$$\delta_\Phi S[\eta,\Phi;\delta\Phi] =$$
$$= \int_{t_0}^{t_1}\int_X\int_{-h}^\eta \left(\partial_t(\delta\Phi) + \nabla_{\boldsymbol{x}}\Phi\cdot\nabla_{\boldsymbol{x}}(\delta\Phi) + \partial_z\Phi\partial_z(\delta\Phi)\right)dz\,d\boldsymbol{x}\,dt +$$
$$+ \int_{t_0}^{t_1}\int_{\partial X_e}\int_{-h}^{\eta_e} V_e(\boldsymbol{x},z,t)[\delta\Phi]_{(\boldsymbol{x},z)\in W_e}\,dz\,d\boldsymbol{x}\,dt,, \tag{A.6}$$

$$\delta_\eta S[\eta,\Phi;\delta\eta] = \int_{t_0}^{t_1}\int_X \left[\partial_t\Phi + \frac{1}{2}|\nabla_{\boldsymbol{x}}\Phi|^2 + \frac{1}{2}(\partial_z\Phi)^2 + gz\right]_{z=\eta}\delta\eta\,d\boldsymbol{x}\,dt. \tag{A.7}$$

Combining Eqs. (A.6), (A.7) with Eqs. (A.4), (A.5), and using the standard assumptions,

(a) isochronality conditions: $[\delta\eta]_{t=t_0,t_1} = 0$ and $[\delta\varphi_m]_{t=t_0,t_1} = 0$, $m\geq -2$,

(b) rest-at-infinity conditions: $|\delta\eta|_{\boldsymbol{x}\in\partial X_\infty} \to 0$ and $|\delta\varphi_m|_{\boldsymbol{x}\in\partial X_\infty} \to 0$, $m\geq -2$ as $|(\boldsymbol{x},z)|_{(\boldsymbol{x},z)\in W_\infty} \to \infty$ and

(c) $[\delta\eta]_{\boldsymbol{x}\in\partial X_e} = 0$ on the excitation boundary, since $[\eta]_{\boldsymbol{x}\in\partial X_e} =$ given there,

we obtain, after copious analytical manipulations, the formulae:

$$\delta_{\varphi_m}\widetilde{S}[\eta,\boldsymbol{\varphi};\delta\varphi_m] =$$
$$= \int_{t_0}^{t_1}\int_X \left\{\left[N_\eta\cdot\left(\nabla_{\boldsymbol{x}}(\boldsymbol{\varphi}^\mathrm{T}\mathbf{Z}),\partial_z(\boldsymbol{\varphi}^\mathrm{T}\mathbf{Z})\right)\right]_{z=\eta} - \partial_t\eta\right] + \sum_{n=-2}^\infty L_{mn}\varphi_n\right\}\delta\varphi_m\,d\boldsymbol{x}\,dt +$$
$$+ \int_{t_0}^{t_1}\int_{\partial X_0}\int_{-h}^\eta n_{\partial X_0}\cdot\left[\nabla_{\boldsymbol{x}}(\boldsymbol{\varphi}^\mathrm{T}\mathbf{Z})Z_m\right]_{(\boldsymbol{x},z)\in W_0}dz\,[\delta\varphi_m]_{\boldsymbol{x}\in\partial X_0}\,d\boldsymbol{x}\,dt + \tag{A.8}$$
$$+ \int_{t_0}^{t_1}\int_{\partial X_e}\int_{-h}^{\eta_e} n_{\partial X_e}\cdot\left[\left(\nabla_{\boldsymbol{x}}(\boldsymbol{\varphi}^\mathrm{T}\mathbf{Z})+V_e\right)Z_m\right]_{(\boldsymbol{x},z)\in W_e}dz\,[\delta\varphi_m]_{\boldsymbol{x}\in\partial X_e}\,d\boldsymbol{x}\,dt,$$

$$\delta_\eta \widetilde{S}[\eta,\boldsymbol{\varphi};\delta\eta] =$$
$$= \int_{t_0}^{t_1} \int_X \left\{ \left[ \partial_t(\boldsymbol{\varphi}^T \mathbf{Z}) + \frac{1}{2}|\nabla_x(\boldsymbol{\varphi}^T \mathbf{Z})|^2 + \frac{1}{2}(\partial_z(\boldsymbol{\varphi}^T \mathbf{Z}))^2 + gz \right]_{z=\eta} \right.$$
$$- \sum_{m=-2}^\infty \left( \sum_{n=-2}^\infty l_{mn} \varphi_n \right) \varphi_m +$$
$$\left. + \left( -\partial_t \eta + N_\eta \cdot \left[ (\nabla_x(\boldsymbol{\varphi}^T \mathbf{Z}), \partial_z(\boldsymbol{\varphi}^T \mathbf{Z})) \right]_{z=\eta} \right) \left[ \boldsymbol{\varphi}^T \partial_\eta \mathbf{Z} \right]_{z=\eta} \right\} \delta\eta \, d\boldsymbol{x} \, dt, \quad \text{(A.9)}$$

where $N_\eta = (-\nabla_x \eta, 1)$, $N_h(-\nabla_x h, -1)$. The symbols $L_{mn}$ and $l_{mn}$, appearing in the right-hand side of Eqs. (A.8) and (A.9), express horizontal differential operators, defined by the equations

$$\sum_{n=-2}^\infty L_{mn} \varphi_n = \sum_{n=-2}^\infty \left( A_{m,n} \nabla_x^2 + \boldsymbol{B}_{mn} \cdot \nabla_x + C_{m,n} \right) \varphi_n, \quad m \geq -2, \quad \text{(A.10)}$$

$$\sum_{n=-2}^\infty l_{mn} \varphi_n = \sum_{n=-2}^\infty \left( a_{m,n} \nabla_x^2 + \boldsymbol{b}_{mn} \cdot \nabla_x + c_{m,n} \right) \varphi_n, \quad m \geq -2, \quad \text{(A.11)}$$

where $A_{m,n}$, $\boldsymbol{B}_{m,n} = (B^1_{m,n}, B^2_{m,n})$ and $C_{m,n}$ are the $(\boldsymbol{x},t)$–dependent coefficients given by Eqs. (18), while the coefficients $a_{m,n}$, $\boldsymbol{b}_{m,n} = (b^1_{m,n}, b^2_{m,n})$ and $c_{m,n}$ are given by the formulae

$$a_{m,n} = \int_{-h}^\eta Z_n \, \partial_\eta Z_m \, dz, \quad \text{(A.12)}$$

$$\boldsymbol{b}_{m,n} = 2\int_{-h}^\eta \nabla_x Z_n \, \partial_\eta Z_m \, dz + \nabla_x h \left[ Z_n \, \partial_\eta Z_m \right]_{z=-h}, \quad \text{(A.13)}$$

$$c_{m,n} = \int_{-h}^\eta \left( \Delta_x Z_n + \partial_z^2 Z_n \right) \partial_\eta Z_m \, dz + (\nabla_x h, 1) \cdot \left[ \begin{pmatrix} \nabla_x Z_n \\ \partial_z Z_n \end{pmatrix} \partial_\eta Z_m \right]_{z=-h}. \quad \text{(A.14)}$$

The variational equation (A.3), in conjunction with Eqs. (A.8) and (A.9), and the arbitrariness of the variations $\delta\eta$ and $\delta\varphi_m$, $m \geq -2$, result in the following Euler-Lagrange equations:

$$\delta\eta: \quad \left[ \partial_t(\boldsymbol{\varphi}^T \mathbf{Z}) + \frac{1}{2}|\nabla_x(\boldsymbol{\varphi}^T \mathbf{Z})|^2 + \frac{1}{2}(\partial_z(\boldsymbol{\varphi}^T \mathbf{Z}))^2 \right]_{z=\eta} +$$
$$+ g\eta - \sum_{m=-2}^\infty \left( \sum_{n=-2}^\infty l_{mn} \varphi_n \right) \varphi_m + \quad \text{(A.15)}$$
$$+ \left( -\partial_t \eta + N_\eta \cdot \left[ (\nabla_x(\boldsymbol{\varphi}^T \mathbf{Z}), \partial_z(\boldsymbol{\varphi}^T \mathbf{Z})) \right]_{z=\eta} \right) \left[ \boldsymbol{\varphi}^T \partial_\eta \mathbf{Z} \right]_{z=\eta} = 0,$$

$$\delta\varphi_m: \quad \partial_t \eta - N_\eta \cdot \left[ \nabla_x (\boldsymbol{\varphi}^T Z) \right]_{z=\eta} + \sum_{n=-2}^{\infty} L_{mn} \varphi_n = 0, \quad m \geq -2, \tag{A.16}$$

complemented with the lateral boundary conditions:

$$\left[ \delta\varphi_m \right]_{x \in \partial X_0}: \quad n_{\partial X_0} \cdot \left[ \sum_{n=-2}^{\infty} \left( \nabla_x \varphi_n A_{m,n} + \frac{1}{2} \varphi_n B_{m,n}^{\text{int}} \right) \right]_{x \in \partial X_0} = 0, \tag{A.17}$$

$$\left[ \delta\varphi_m \right]_{x \in \partial X_e}: \quad n_{\partial X_e} \cdot \left[ \sum_{n=-2}^{\infty} \left( \nabla_x \varphi_n A_{m,n} + \frac{1}{2} \varphi_n B_{m,n}^{\text{int}} \right) \right]_{x \in \partial X_e} = -g_m, \tag{A.18}$$

where $g_m = \int_{-h}^{\eta_e} V_e(x, z, t) [Z_m]_{(x,z) \in W_e} dz$, and $B_{m,n}^{\text{int}}$ is defined by Eq. (45a). The Euler-Lagrange equations (A.15) and (A.16) are essentially the same with the ones obtained by [47] and [48]. Although these equations can be (and have been) used for numerical simulations by truncating the two infinite series $\sum_{n=-2}^{\infty} L_{mn} \varphi_n$ and $\sum_{n=-2}^{\infty} l_{mn} \varphi_n$, it is possible to reformulate them in a much simpler, yet equivalent, form. The main steps leading to this simplification are briefly described herein, referring to [50] for detailed proofs. First, it is proved that, if $\boldsymbol{\varphi} = \{\varphi_n(x;t)\}_{n=-2}^{\infty}$ satisfies the system of Eqs. (A.16) at every $t \in [t_0, t_1]$, then, it satisfies the equations

$$\partial_t \eta - N_\eta \cdot \left[ \nabla (\boldsymbol{\varphi}^T Z) \right]_{z=\eta} = 0, \quad x \in X, \tag{A.19}$$

$$\sum_{n=-2}^{\infty} L_{mn} \varphi_n = 0, \quad m \geq -2, \quad x \in X. \tag{A.20}$$

That is, Eqs. (A.16) can be separated in a simple evolution equation, Eq. (A.19), which is the kinematic free-surface condition, and a system of time-independent (yet parametrically dependent on time through the coefficients of $L_{mn}$) horizontal differential equations (A.20), which is identical with Eqs. (17a) of Sec. 2. Further, it can be proved that

$$\sum_{n=-2}^{\infty} \ell_{mn}[\eta, h] \varphi_n = 0, \quad m \geq -2,$$

which, in conjunction with Eq. (A.19), simplifies Eq. (A.15) to

$$\left[ \partial_t (\boldsymbol{\varphi}^T Z) + \frac{1}{2} |\nabla_x (\boldsymbol{\varphi}^T Z)|^2 + \frac{1}{2} (\partial_z (\boldsymbol{\varphi}^T Z))^2 \right]_{z=\eta} + g\eta = 0. \tag{A.21}$$

Now, setting $\psi = [\boldsymbol{\varphi}^T Z]_{z=\eta}$ and recalling that $[Z_n]_{z=\eta} = 1$ by construction (see Eq. (7)), we get $\psi = \sum_{n=-2}^{\infty} \varphi_n$, which is the same as Eq. (17b). Finally, using the equations

$$\left[ \partial_t (\boldsymbol{\varphi}^T Z) \right]_{z=\eta} = \partial_t \psi - \boldsymbol{\varphi}^T [\partial_z Z]_{z=\eta} \partial_t \eta,$$

$$\left[ \nabla_x (\boldsymbol{\varphi}^T Z) \right]_{z=\eta} = \nabla_x \psi - \left( \boldsymbol{\varphi}^T [\partial_z Z]_{z=\eta} \right) \nabla_x \eta,$$

$$\left[\partial_z(\boldsymbol{\varphi}^T \boldsymbol{Z})\right]_{z=\eta} = \boldsymbol{\varphi}^T \left[\partial_z \boldsymbol{Z}\right]_{z=\eta} = h_0^{-1} \varphi_{-2} + \mu_0 \psi,$$

which relate the derivatives $\left[\partial_\alpha(\boldsymbol{\varphi}^T \boldsymbol{Z})\right]_{z=\eta}$ with the derivatives $\partial_\alpha \psi \equiv \partial_\alpha \left[\boldsymbol{\varphi}^T \boldsymbol{Z}\right]_{z=\eta}$, $\alpha \in \{t, x_1, x_2, z\}$ (for the proofs see [50] ([8])), we see that Eqs. (A.19) and (A.21) become identical with Eqs. (16b) of Sec. 2. Thus, the (sketch of the) proof of the HCMS, Eqs. (16), (17) and (18), has been completed.

## Appendix B: Calculation of the basic vertical integrals

In this appendix analytic expressions for the basic integrals, involved in the calculations of the coefficients $A_{m,n}$, $\boldsymbol{B}_{m,n}$, $C_{m,n}$, $m,n \geq -2$, Eqs. (47)-(49), are presented. For their calculation, integration by parts is utilized and, therefore, the boundary values of the eigenfunctions $Z_n$ and the auxiliary functions $W_n$ on the free-surface, $z = \eta$, and the bottom, $z = -h$, are needed. These boundary values are only dependent on the local depth $H = \eta + h$ and the eigenvalues $k_n$, $n \geq 0$. The exact boundary values of the eigenfunctions $Z_n$ are as follows: At the free surface, $z = \eta$, we have $\left[Z_n\right]_\eta = 1$ for all $n \geq -2$, by construction (see Eq. (7)). At the bottom, $z = -h$, we have

$$\left[Z_n\right]_{-h} = a_n^{(0)}, \qquad n = -2, -1, \tag{B.1a}$$

$$\left[Z_n\right]_{-h} = \begin{cases} k_0^{-1} \sqrt{k_0^2 - \mu_0^2}, & n = 0, \\ (-1)^n k_n^{-1} \sqrt{k_n^2 + \mu_0^2}, & n \geq 1. \end{cases} \tag{B.1b}$$

Note that Eqs. (B.1) are also needed for the calculation of the boundary terms, $\boldsymbol{B}_{m,n}^b$ and $C_{m,n}^b$, of the coefficients; see Eqs. (45) and (46). The boundary values of the auxiliary functions $W_n$, $n \geq 0$, Eqs. (29a,b), are given by the formulae

$$\left[W_n\right]_\eta = \begin{cases} \mu_0 k_0^{-1}, & n = 0 \\ -\mu_0 k_n^{-1}, & n \geq 1 \end{cases}, \qquad \text{and} \qquad \left[W_n\right]_{-h} = 0, \qquad n \geq 0. \tag{B.2a,b}$$

For the systematic presentation of the results to follow, use is also made of the vertical integrals $J(s; W_m)$, even if they do not explicitly appear in the formulae (47) – (49) of the coefficients $A_{m,n}$, $\boldsymbol{B}_{m,n}$, $C_{m,n}$, $m, n \geq -2$.

---

([8]) Especially the third equation is strongly dependent on the specific choice of the vertical functions $\{Z_n\}_{n=-2}^\infty$ made herein.

**(a) Vertical integrals of the form $J(s;Z_m)$ and $J(s;W_m)$**

For any exponent $s \geq 0$, the integrals $J(s;Z_m)$, $m = -2, -1$, are computed straightforwardly. The results read as follows:

$$J(s;Z_m) = a_m^{(2)} \frac{H^{s+3}}{s+3} + a_m^{(1)} \frac{H^{s+2}}{s+2} + a_m^{(0)} \frac{H^{s+1}}{s+1}, \quad \begin{cases} m = -2, -1, \\ s = 0, 1, 2, \dots, \end{cases} \quad \text{(B.3)}$$

where $H = \eta + h$, and the coefficients $a_m^{(0)}, a_m^{(1)}, a_m^{(2)}$ are given by Eqs. (11a,b). For $m \geq 0$, the relations

$$Z_m = \partial_z W_m / k_m, \quad m \geq 0 \quad \text{and} \quad \begin{cases} W_0 = \partial_z Z_0 / k_0, \\ W_m = -\partial_z Z_m / k_m, \quad m \geq 1, \end{cases}$$

permit us to apply integration by parts, obtaining the following recursive relations, for the integrals $J(s;Z_m)$ and $J(s;W_m)$, with respect to the exponent $s$:

$$J(s;Z_m) = \begin{cases} H^s \dfrac{\mu_0}{k_0^2} - \dfrac{s}{k_0} J(s-1; W_0), & m = 0, \\[2mm] -H^s \dfrac{\mu_0}{k_m^2} - \dfrac{s}{k_m} J(s-1; W_m), & m \geq 1, \end{cases} \quad s = 1, 2, \dots, \quad \text{(B.4)}$$

$$J(s;W_m) = \begin{cases} \dfrac{H^s}{k_0} - \dfrac{s}{k_0} J(s-1; Z_0), & m = 0, \\[2mm] -\dfrac{H^s}{k_m} + \dfrac{s}{k_m} J(s-1; Z_m), & m \geq 1. \end{cases} \quad s = 1, 2, \dots \quad \text{(B.5)}$$

In deriving the above relations, use is also made of the local dispersion relations (10a,b), to eliminate the hyperbolic (trigonometric) function $\tanh(k_0 H)$ ($\tan(k_m H)$). Initializations of the recursive relations (B.4) and (B.5) are provided by the following formulae, which are easily obtained by direct calculations:

$$J(0; Z_0) = \frac{\mu_0}{k_0^2}, \qquad J(0; Z_m) = -\frac{\mu_0}{k_m^2}, \qquad m \geq 1, \quad \text{(B.6)}$$

and

$$J(0; W_m) = \begin{cases} k_0^{-1} - k_0^{-2} \sqrt{k_0^2 - \mu_0^2}, & m = 0, \\[1mm] -k_m^{-1} + (-1)^m k_m^{-2} \sqrt{k_m^2 + \mu_0^2}, & m \geq 1. \end{cases} \quad \text{(B.7)}$$

Eqs. (B.3) - (B.7) suffice for the calculation of the coefficients $B_{m,n}$ and $C_{m,n}$ for $n = -2$, $-1$ and $m \geq -2$.

**(b) Vertical integrals of the form** $J(s;Z_n Z_m)$ **for** $s = 0, 1, 2$

For the case $n = -2, -1$ and $m \geq -2$, we can easily derive the required formulae through straightforward integration:

$$J(s; Z_n Z_m) = \sum_{k=0}^{2} a_n^{(k)} J(s+k; Z_m), \quad \begin{cases} n = -2, -1, \\ m \geq -2, \end{cases} \quad s = 0, 1, 2, \ldots. \quad \text{(B.8)}$$

Notice that Eq. (B.8) holds also for $n \geq -2$ and $m = -2, -1$, due to the symmetry of the vertical integrals $J(s; Z_n Z_m)$ with respect to indexes $m, n$.

For $m, n \geq 0$, we once again invoke integration by parts and utilize the normalization condition Eq. (7), obtaining the following recursive relations for the calculation of these integrals, for $s \geq 1$

$$J(s; Z_n Z_m) = \begin{cases} \dfrac{H^{s+1}}{2(s+1)}\left(1 - \dfrac{\mu_0^2}{k_0^2}\right) + \dfrac{H^s \mu_0}{2k_0^2} - \dfrac{s}{2k_0} J(s-1; W_0 Z_0), & m = n = 0, \\[2mm] \dfrac{H^{s+1}}{2(s+1)}\left(1 + \dfrac{\mu_0^2}{k_m^2}\right) - \dfrac{H^s \mu_0}{2k_m^2} - \dfrac{s}{2k_m} J(s-1; W_m Z_m), & m = n \geq 1, \quad \text{(B.9)} \\[2mm] s \, \dfrac{k_n J(s-1; W_n Z_m) - k_m J(s-1; W_m Z_n)}{k_m^2 - k_n^2}, & m, n \geq 0, \; m \neq n. \end{cases}$$

For $s = 0$ we find, by direct integration, the formulae

$$J(0; Z_m Z_m) = \begin{cases} \dfrac{H}{2}\left(1 - \dfrac{\mu_0^2}{k_0^2}\right) + \dfrac{\mu_0}{2k_0^2}, & m = n = 0, \\[2mm] \dfrac{H}{2}\left(1 + \dfrac{\mu_0^2}{k_m^2}\right) - \dfrac{\mu_0}{2k_m^2}, & m = n \geq 1, \end{cases} \quad \text{(B.10a)}$$

$$J(0; Z_n Z_m) = 0, \quad m, n \geq 0, \; m \neq n. \quad \text{(B.10b)}$$

For the exploitation of the recursive relations (B.9) we need the calculation of the vertical integrals $J(s; W_n Z_m)$, which is presented below.

**(c) Vertical integrals of the form** $J(s; W_n Z_m)$ **for** $s = 0, 1$

For the case $m = -2, -1$ and $n \geq -2$, we easily derive the required formulae, through straightforward integration:

$$J(s; W_n Z_m) = \sum_{k=0}^{2} a_m^{(k)} J(s+k; W_n), \quad \begin{cases} m = -2, -1, \\ n \geq 0, \end{cases} \quad s = 0, 1. \quad \text{(B.11)}$$

For the cases $m = n \geq 0$ we can utilize integration by parts, deriving the formulae (needed only for $s = 0, 1$)

$$J(s; W_n Z_m) = \begin{cases} \dfrac{1}{2k_0}\left(H^s - \left[(z+h)^s\right]_{-h}\dfrac{k_0^2 - \mu_0^2}{k_0^2}\right) - \dfrac{s}{2k_0}J(0; Z_0 Z_0), & m = n = 0, \\ -\dfrac{1}{2k_m}\left(H^s - \left[(z+h)^s\right]_{-h}\dfrac{k_m^2 + \mu_0^2}{k_m^2}\right) + \dfrac{s}{2k_m}J(0; Z_m Z_m), & m = n \geq 1. \end{cases}$$

(B.12)

For the case $m \geq 1$, $n = 0$ and $s = 0, 1$, we can derive the following relations

$$J(s; W_0 Z_m) = \begin{cases} \dfrac{-H\mu_0^2 k_0^{-1} + k_0 H + \mu_0 k_0^{-1}}{k_m^2 + k_0^2}, & s = 1, \\ \dfrac{-\mu_0^2 k_0^{-1} + k_0 - (-1)^m k_m^{-1}\sqrt{k_0^2 - \mu_0^2}\sqrt{k_m^2 + \mu_0^2}}{k_m^2 + k_0^2}, & s = 0, \end{cases} \quad m \geq 1. \quad (B.13)$$

For the case $m = 0$ and $n \geq 1$ we have the results

$$J(s; W_n Z_0) = \begin{cases} \dfrac{-H\mu_0^2 k_n^{-1} - k_n H + \mu_0 k_n^{-1}}{k_n^2 + k_0^2}, & s = 1, \\ \dfrac{-\mu_0^2 k_n^{-1} - k_n + (-1)^n k_0^{-1}\sqrt{k_0^2 - \mu_0^2}\sqrt{k_n^2 + \mu_0^2}}{k_n^2 + k_0^2}, & s = 0. \end{cases} \quad n \geq 1. \quad (B.14)$$

For the case $m, n \geq 1$ with $m \neq n$ we have

$$J(s; W_n Z_m) = \begin{cases} -\dfrac{H\mu_0^2 k_n^{-1} + k_n H - \mu_0 k_n^{-1}}{k_n^2 - k_m^2}, & s = 1, \\ -\dfrac{\mu_0^2 k_n^{-1} + k_n - (-1)^{m+n} k_m^{-1}\sqrt{k_m^2 + \mu_0^2}\sqrt{k_n^2 + \mu_0^2}}{k_n^2 - k_m^2}, & s = 0. \end{cases}$$

(B.15)

## Appendix C. Fast and accurate calculation of local wavenumbers $k_n(\boldsymbol{x}, t)$

We present here a method of computation of the local wavenumbers $k_n(\boldsymbol{x}, t)$, $n \geq 0$, through a semi-explicit solution algorithm of the transcendental equations (10a,b), repeated here in the form

$$\mu - \kappa_0 \tanh(\kappa_0) = 0, \qquad \mu + \kappa_n \tan(\kappa_n) = 0, \; n \geq 1, \qquad \text{(C.1a,b)}$$

with $\mu = \mu(\boldsymbol{x}, t) = \mu_0 H(\boldsymbol{x}, t)$ and $\kappa_n = \kappa_n(\boldsymbol{x}, t) = k_n(\boldsymbol{x}, t) H(\boldsymbol{x}, t)$.

For the case $n = 0$, Eq. (C.1a) is solved by the Newton-Raphson (NR) method, initialized by using an available explicit approximation, accurate up to the third or fourth decimal digit, for all values of the parameter $\mu$ [81]:

$$^0\kappa_0(\mu) = \frac{\mu + \mu^{1.986}\exp(-1.863 + 1.198\,\mu^{1.366})}{\sqrt{\tanh\mu}}. \tag{C.2}$$

Formula (C.2) has maximum relative error of approximately $10^{-4}$, and leads to machine accuracy solutions, for any value of $\mu = \mu_0 H(\mathbf{x},t)$, within 2 NR iterations.

For the case of $n \geq 1$, we adopt a recently developed iterative procedure [62], valid for all values of $\mu = \mu_0 H(\mathbf{x},t)$:

$$^{j+1}\kappa_n = n\pi + \frac{\mu\left(n\pi - {}^j\kappa_n\right)}{\mu^2 + {}^j\kappa_n^2 - \mu} - \frac{\mu^2 + {}^j\kappa_n^2}{\mu^2 + {}^j\kappa_n^2 - \mu}\,\mathrm{Arctan}\left(\frac{\mu}{{}^j\kappa_n}\right). \tag{C.3}$$

Using the formula

$$^0\kappa_n = n\pi - \left[\frac{\mu^2 + n^2\pi^2 - \mu}{\mu^2 + n^2\pi^2 - 2\mu}\right]\mathrm{Arctan}\left(\frac{\mu}{n\pi}\right), \tag{C.4}$$

to initialize (C.3), machine-accuracy solutions are obtained with no more than 3 iterations. More details can be found in [62].

**Appendix D: Finite-difference linear system for the substrate problem Eqs. (17a,b)**

We present here the linear algebraic system resulting from the fourth-order FD discretization of the substrate system (17a,b), with vertical wall conditions at the lateral boundaries $x = a$ ($i = 1$) and $x = b$ ($i = N_X$); see Eq. (A.17). Using the notation $\psi^i = \psi(x_i, t)$, $A_{m,n}^i = A_{m,n}(\eta(x_i,t), h(x_i))$, and similarly for $B_{m,n}^i$ and $C_{m,n}^i$, the aforementioned linear system is written in the form

$$\sum_{n=-2}^{M}\left(-\frac{25 A_{m,n}^1}{12\Delta x} + \frac{B_{m,n}^1}{2}\right)\varphi_n^1 + \frac{A_{m,n}^1}{\Delta x}\left(4\varphi_n^2 - 3\varphi_n^3 + \frac{4}{3}\varphi_n^4 - \frac{1}{4}\varphi_n^5\right) = 0, \quad (i=1), \tag{D.1}$$

$$\sum_{n=-2}^{M}\left(\frac{5A_{m,n}^2}{6\Delta x^2} - \frac{B_{m,n}^2}{4\Delta x}\right)\varphi_n^1 - \left(\frac{5A_{m,n}^2}{4\Delta x^2} - \frac{5B_{m,n}^2}{6\Delta x} - C_{mn}^2\right)\varphi_n^2 - \left(\frac{A_{m,n}^2}{3\Delta x^2} - \frac{3B_{m,n}^2}{2\Delta x}\right)\varphi_n^3$$
$$+ \left(\frac{7A_{m,n}^2}{6\Delta x^2} - \frac{B_{m,n}^2}{2\Delta x}\right)\varphi_n^4 - \left(\frac{A_{m,n}^2}{2\Delta x^2} - \frac{B_{m,n}^2}{12\Delta x}\right)\varphi_n^5 + \frac{A_{m,n}^2}{12\Delta x^2}\varphi_n^6 = 0, \quad (i=2), \tag{D.2}$$

$$\sum_{n=-2}^{M} \left( -\frac{A_{m,n}^i}{12\Delta x^2} + \frac{B_{m,n}^i}{12\Delta x} \right) \varphi_n^{i-2} + \left( \frac{4A_{m,n}^i}{3\Delta x^2} - \frac{2B_{m,n}^i}{3\Delta x} \right) \varphi_n^{i-1} - \left( \frac{5A_{m,n}^i}{2\Delta x^2} - C_{mn}^i \right) \varphi_n^i$$
$$+ \left( \frac{4A_{m,n}^i}{3\Delta x^2} + \frac{2B_{m,n}^i}{3\Delta x} \right) \varphi_n^{i+1} - \left( \frac{A_{m,n}^i}{12\Delta x^2} + \frac{B_{m,n}^i}{12\Delta x} \right) \varphi_n^{i+2} = 0, \quad i=3,\dots,N_X-2, \quad (D.3)$$

$$\sum_{n=-2}^{M} \left( \frac{5A_{m,n}^{N_X-1}}{6\Delta x^2} + \frac{B_{m,n}^{N_X-1}}{4\Delta x} \right) \varphi_n^{N_X} - \left( \frac{5A_{m,n}^{N_X-1}}{4\Delta x^2} + \frac{5B_{m,n}^{N_X-1}}{6\Delta x} - C_{m,n}^{N_X-1} \right) \varphi_n^{N_X-1} - \left( \frac{A_{m,n}^{N_X-1}}{3\Delta x^2} + \frac{3B_{m,n}^{N_X-1}}{2\Delta x} \right) \varphi_n^{N_X-2}$$
$$+ \left( \frac{7A_{m,n}^{N_X-1}}{6\Delta x^2} + \frac{B_{m,n}^{N_X-1}}{2\Delta x} \right) \varphi_n^{N_X-3} - \left( \frac{A_{m,n}^{N_X-1}}{2\Delta x^2} + \frac{B_{m,n}^{N_X-1}}{12\Delta x} \right) \varphi_n^{N_X-4} + \frac{A_{m,n}^{N_X-1}}{12\Delta x^2} \varphi_n^{N_X-5} = 0, \quad (i = N_X - 1),$$
$$(D.4)$$

$$\sum_{n=-2}^{M} \left( \frac{25 A_{m,n}^{N_X}}{12\Delta x} + \frac{B_{m,n}^{N_X}}{2} \right) \varphi_n^{N_X} - \frac{A_{m,n}^{N_X}}{\Delta x} \left( 4\varphi_n^{N_X-1} - 3\varphi_n^{N_X-2} + \frac{4}{3}\varphi_n^{N_X-3} - \frac{1}{4}\varphi_n^{N_X-4} \right) = 0, \quad (D.5)$$
$$(i = N_X),$$

$$\sum_{n=-2}^{M} \varphi_n^i = \psi^i, \quad i=1,\dots,N_X \quad (D.6)$$

with $m = -2, \dots, M-1$.

### Appendix E. Description of the 3D bathymetry

In order to describe exactly the 3D bathymetry of the case study presented in Sec. 6.5, we introduce the auxiliary (smooth-step) functions

$$\begin{cases} E_+(x_i; a, b) = \frac{1}{2}\left[ 1 + \tanh\left[ b(x_i - a) \right] \right], \\ E_-(x_i; a, b) = \frac{1}{2}\left[ 1 - \tanh\left[ b(x_i - a) \right] \right], \quad (i=1,2) \\ E\left( x_i; \begin{array}{c} a_1, b_1 \\ a_2, b_2 \end{array} \right) = E_+(x_i; a_1, b_1) E_-(x_i; a_2, b_2). \end{cases} \quad (E.1)$$

The function $E_+$ permits us to describe the gradual deepening, along the $x_1$-direction, from $h_0 = 1m$ to $h_{tr} = 3m$, by means of the formula (in meters)

$$h_1(x) = 1 + 2 E_+\left( x_1; a_1 = 75, b_1 = 0.1 \right). \quad (E.2)$$

The function $E$ is used to construct a system of banks and trenches, as shown in the right part of Fig. 15, with minimum depth $h_0 = 1m$ and maximum depth $h_{tr} = 3m$, extending parallel to the $x_1$-direction, by means of the formula (in meters)

$$h_2(\boldsymbol{x}) = 3 - 2E\left(x_2; \begin{matrix} 6 & 0.75 \\ 18 & 0.75 \end{matrix}\right) - 2E\left(x_2; \begin{matrix} 32 & 0.75 \\ 44 & 0.75 \end{matrix}\right). \tag{E.3}$$

The 3D bathymetry, shown in Fig. 15, is set to be

$$h(\boldsymbol{x}) = \min\{h_1(\boldsymbol{x}), h_2(\boldsymbol{x})\}.$$